\documentclass[8pt,preprint]{aastex}
\usepackage[]{aastexug,natbib,epsfig,graphicx,rotating,lscape,rotfloat,subfigure,lineno,textcomp}
\def\arcsec{$^{\prime\prime}$}
\def\arcmin{$^{\prime}$}
\newcommand{\NtwoH}{N$_{2}$H$^{+}$}
\newcommand{\HCO}{HCO$^{+}$}
\newcommand{\vlsr}{V$_{\mathrm{lsr}}$}

\def\deg{$^{\circ}$}
\newcommand{\Msun}{\mbox{${M}_{\sun}$}}
\newcommand{\Jybm}{Jy~beam$^{-1}$}

\newcommand{\Jybmkms}{Jy~beam$^{-1}$~km~s$^{-1}$}
\newcommand{\kms}{km~s$^{-1}$}
\newcommand{\kmsppc}{km~s$^{-1}$~pc$^{-1}$}

\begin{document}

\title{CARMA LARGE AREA STAR FORMATION SURVEY: PROJECT OVERVIEW WITH ANALYSIS OF DENSE GAS STRUCTURE AND KINEMATICS IN BARNARD 1}

\shorttitle{CARMA Large Area Star Formation Survey: Barnard~1}

\author{
Shaye Storm\altaffilmark{1}, 
Lee G. Mundy\altaffilmark{1},
Manuel Fern\'{a}ndez-L\'{o}pez\altaffilmark{2,3},
Katherine I. Lee\altaffilmark{2,1},
Leslie W. Looney\altaffilmark{2},
Peter Teuben\altaffilmark{1},
Erik W. Rosolowsky\altaffilmark{4,5},
H\'{e}ctor G. Arce\altaffilmark{6},
Eve C. Ostriker\altaffilmark{7},
Dominique M. Segura-Cox\altaffilmark{2},
Marc W. Pound\altaffilmark{1},
Demerese M. Salter\altaffilmark{1},
Nikolaus H. Volgenau\altaffilmark{8},
Yancy L. Shirley\altaffilmark{9},
Che-Yu Chen\altaffilmark{1},
Hao Gong\altaffilmark{1},
Adele L. Plunkett\altaffilmark{6},
John J. Tobin\altaffilmark{10},
Woojin Kwon\altaffilmark{11},
Andrea Isella\altaffilmark{12},
Jens Kauffmann\altaffilmark{13},
Konstantinos Tassis\altaffilmark{14,15},
Richard M. Crutcher\altaffilmark{2},
Charles F. Gammie\altaffilmark{2},
Leonardo Testi\altaffilmark{16}
}
\altaffiltext{1}{Department of Astronomy, University of Maryland, College Park, MD 20742, USA; sstorm@astro.umd.edu}   
\altaffiltext{2}{Department of Astronomy, University of Illinois at Urbana--Champaign, 1002 West Green Street, Urbana, IL 61801, USA}
\altaffiltext{3}{Instituto Argentino de Radioastronom{\'{\i}}a, CCT-La Plata (CONICET), C.C.5, 1894, Villa Elisa, Argentina}
\altaffiltext{4}{University of British Columbia, Okanagan Campus, Departments of Physics and Statistics, 3333 University Way, Kelowna BC V1V 1V7, Canada}
\altaffiltext{5}{University of Alberta, Department of Physics, 4-181 CCIS, Edmonton AB T6G 2E1, Canada}
\altaffiltext{6}{Department of Astronomy, Yale University, P.O.~Box 208101, New Haven, CT 06520-8101, USA}   
\altaffiltext{7}{Department of Astrophysical Sciences, Princeton University, Princeton, NJ 08544, USA} 
\altaffiltext{8}{Owens Valley Radio Observatory, MC 105-24 OVRO, Pasadena, CA 91125, USA}
\altaffiltext{9}{Steward Observatory, 933 North Cherry Avenue, Tucson, AZ 85721, USA}
\altaffiltext{10}{National Radio Astronomy Observatory, Charlottesville, VA 22903, USA}
\altaffiltext{11}{SRON Netherlands Institute for Space Research, Landleven 12, 9747 AD Groningen, The Netherlands}
\altaffiltext{12}{Astronomy Department, California Institute of Technology, 1200 East California Blvd., Pasadena, CA 91125, USA}   
\altaffiltext{13}{Max Planck Institut f\"{u}r Radioastronomie, Auf dem H\"{u}gel 69 D–53121, Bonn Germany}
\altaffiltext{14}{Department of Physics and Institute of Theoretical \& Computational Physics, University of Crete, PO Box 2208, GR-710 03, Heraklion, Crete, Greece}
\altaffiltext{15}{Foundation for Research and Technology - Hellas, IESL, Voutes, 7110 Heraklion, Greece}
\altaffiltext{16}{ESO, Karl-Schwarzschild-Strasse 2 D-85748 Garching bei M\"{u}nchen, Germany}

\email{Accepted to The Astrophysical Journal (ApJ); August 23, 2014}
\vspace{-0.2cm}
\email{(see published version for full-resolution figures)}

\begin{abstract}
We present details of the CARMA Large Area Star Formation Survey
(CLASSy), while focusing on observations of Barnard~1.  CLASSy is a
CARMA Key Project that spectrally imaged \NtwoH, \HCO{}, and HCN
($J=1\rightarrow0$ transitions) across over 800 square arcminutes of
the Perseus and Serpens Molecular Clouds. The observations have
angular resolution near 7\arcsec{} and spectral resolution near
0.16~\kms.  We imaged $\sim$150 square arcminutes of Barnard~1,
focusing on the main core, and the B1 Ridge and clumps to its
southwest. \NtwoH{} shows the strongest emission, with morphology
similar to cool dust in the region, while \HCO{} and HCN trace several
molecular outflows from a collection of protostars in the main core.
We identify a range of kinematic complexity, with \NtwoH{} velocity
dispersions ranging from $\sim$0.05-0.50~\kms{} across the
field. Simultaneous continuum mapping at 3~mm reveals six compact
object detections, three of which are new detections.  A new,
non-binary dendrogram algorithm is used to analyze dense gas
structures in the \NtwoH{} position-position-velocity (PPV) cube. The
projected sizes of dendrogram-identified structures range from about
0.01-0.34~pc.  Size-linewidth relations using those structures show
that non-thermal line-of-sight velocity dispersion varies weakly with
projected size, while rms variation in the centroid velocity rises
steeply with projected size.  Comparing these relations, we propose
that all dense gas structures in Barnard 1 have comparable depths into
the sky, around 0.1-0.2~pc; this suggests that over-dense,
parsec-scale regions within molecular clouds are better described as
flattened structures rather than spherical collections of gas.
Science-ready PPV cubes for Barnard~1 molecular emission are available
for download.
\end{abstract}

\section{INTRODUCTION}
        
Star formation occurs over a wide range of
spatial scales and physical densities in molecular
clouds. Observations reveal that molecular cloud complexes extend for
several tens of parsecs, with low-density molecular gas
($n\le$10$^{2-3}$~cm$^{-3}$) at all spatial scales. Overdensities in
the low-density gas create zones of active star formation at parsec
scales with higher-density gas. These overdensities evolve to even
higher densities ($n\ge$10$^{5-7}$~cm$^{-3}$) as they form pre-stellar
and proto-stellar cores on scales of $\sim$0.01-0.1~pc.  The formation
of initial overdensities and their subsequent evolution to even higher
densities is the focus of current research; supersonic turbulent
flows, magnetic fields, gravitational instabilities, and
thermal-chemical processes are all involved \citep[e.g. reviews
  by][]{2004RvMP...76..125M, 2007prpl.conf...63B, 2007ARAA..45..565M,
  2007prpl.conf...17D, 2007ARA&A..45..339B,
  2012ARA&A..50...29C,2014arXiv1404.2024L}, but the interplay between
them across a broad range of size scales, and across different
environments, is not fully understood.

Large area, high angular resolution surveys of dust and gas from
parsec to thousand AU scales are needed to improve our understanding
of the internal physical state of molecular clouds, and to follow the
structure and kinematics of overdensities on the pathway to forming
stars. The long term goals of these surveys are to understand what
drives and controls the rate of star formation, and to answer why the
stellar initial mass function has its shape and whether that shape
depends on environment. We carried out a Key Project with the Combined
Array for Research in Millimeter-wave Astronomy (CARMA) to address the
need for large area, high resolution gas surveys. The \textbf{C}ARMA
\textbf{L}arge \textbf{A}rea \textbf{S}tar Formation
\textbf{S}urve\textbf{y} (CLASSy) spent 700 hours imaging five large
areas of star formation in the \NtwoH, \HCO, and HCN
($J=1\rightarrow0$) molecular lines. We chose these molecules because
they are dense gas tracers, and potentially sample a range of
chemical/physical environments within a cloud. They are all also
observable in a single CARMA correlator setting.

\begin{figure}[H]
\centering \includegraphics[scale=0.25]{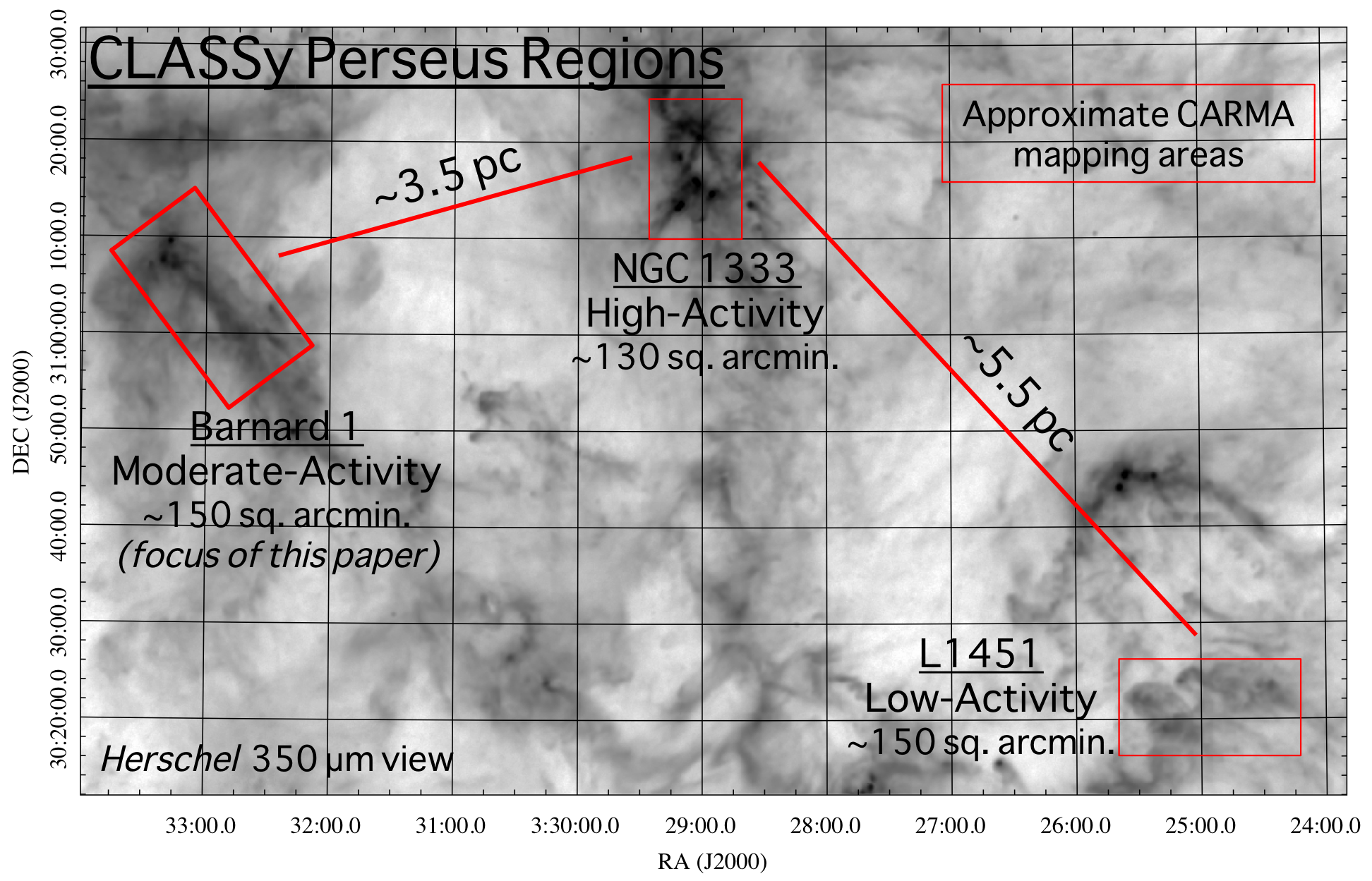}
\caption{\small Overview of the CLASSy regions in the western portion
  of the Perseus Molecular Cloud, as seen by \textit{Herschel}
  350~$\mu$m \citep{2010AA...518L.102A}. The approximate areas mapped
  by CARMA are represented with rectangles (colored red in the online
  version); the actual areas are not single rectangular fields for
  Barnard~1 and L1451. Projected distances between regions are
  given. This paper focuses on dense gas emission in Barnard 1, and
  there will be a separate paper for each region. We assume a distance
  of 235~pc; see text.}
\label{fig:overview}
\end{figure}

\begin{figure}[H]
\centering \includegraphics[scale=0.20]{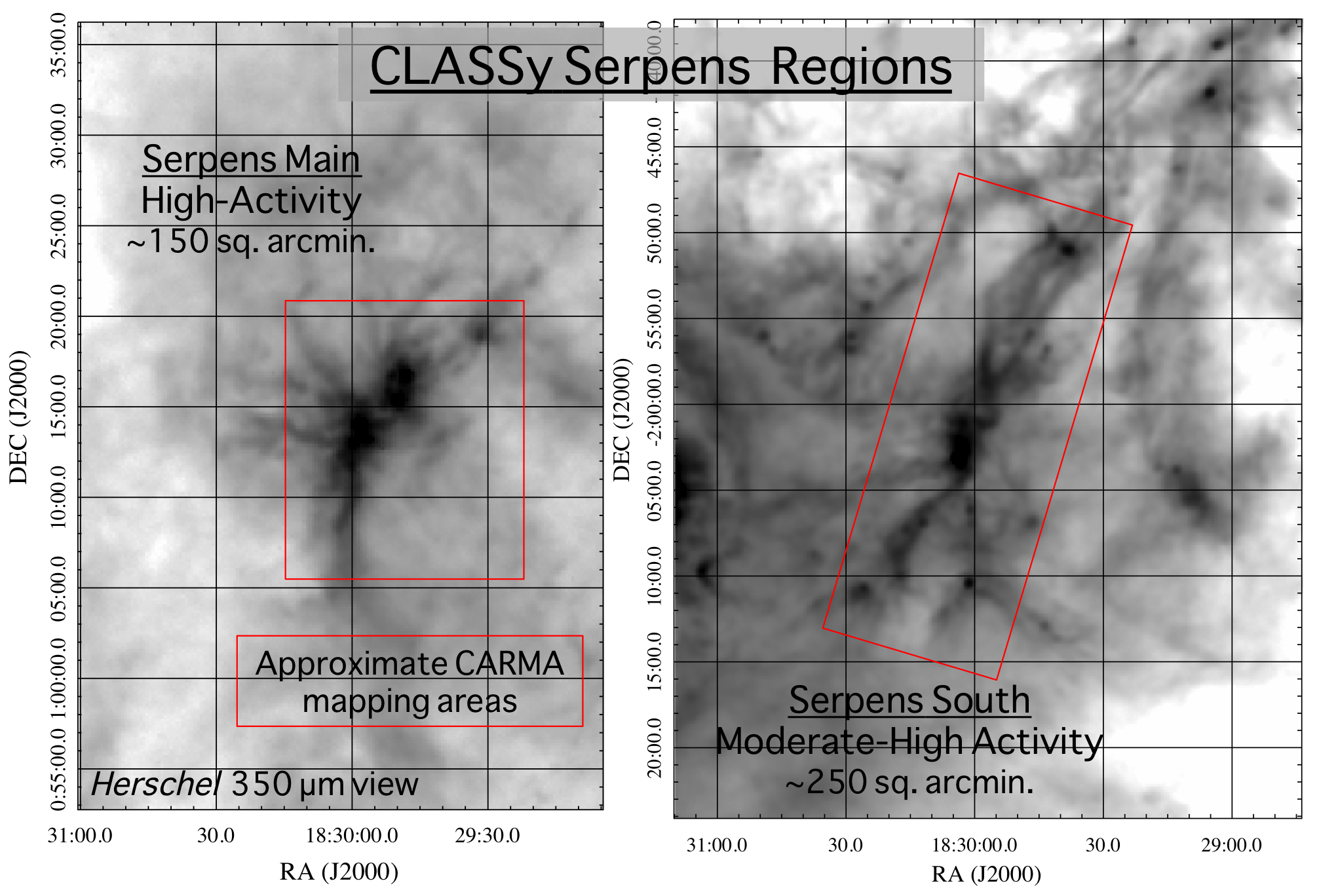}
\caption{\small Overview of CLASSy regions in the Serpens Molecular
  Cloud, as seen by \textit{Herschel} 350~$\mu$m
  \citep{2010AA...518L.102A}. The approximate areas mapped by CARMA
  are represented with rectangles (colored red in the online version);
  the actual areas are not single rectangular fields for either
  region. The Serpens Main data is presented in Lee at al. (2014,
    submitted), while an analysis of the Serpens South \NtwoH{}
    filaments is in \citet{2014ApJ...790L..19F}.}
\label{fig:overviews}
\end{figure}

CLASSy observed the Perseus Molecular Cloud Complex towards NGC~1333,
Barnard 1, and L1451 (see Figure~\ref{fig:overview}), and the Serpens
Molecular Cloud towards Serpens Main and Serpens South (see
Figure~\ref{fig:overviews}), to capture a range of star forming
environments and evolutionary stages. We mapped each region with
7\arcsec{} angular resolution to resolve individual star-forming
cores, and with large enough area to capture the environment in which
the cores formed. Our survey data can be combined with many ancillary
datasets---\textit{Spitzer} catalogs of young stellar objects (YSOs),
\textit{Herschel} images of heating sources, \textit{Herschel} and the
JCMT maps of dust and continuum sources, to name a few---to provide a
total picture of star forming objects and the structure and kinematics
of the dense gas which is fueling their formation.

This paper presents the initial results for the Barnard~1
region. Barnard~1 is located 3.5 pc to the east of NGC 1333 in the
western half of the Perseus Molecular Cloud. We assume a distance of
235~pc based on VLBI parallax measurements towards nearby sources in
Perseus (SVS-13 in NGC~1333 \citep{2008PASJ...60...37H} and L1448-C in
L1448 \citep{2011PASJ...63....1H}).  In the context of the CLASSy
campaign, we classify Barnard~1 as a moderate-activity star forming
region, compared to the highly active NGC~1333 region that is
dominated by a complex array of protostellar outflows, and the
low-activity L1451 complex that has only one identified protostar
\citep{2011ApJ...743..201P}.

Previous surveys of the young stellar content and dust in Barnard~1
reveal a northeast-southwest oriented filament exhibiting a wide range
of star formation (see Figure~\ref{fig:mosaicpt})---progressing from a
tight collection of pre- and proto-stellar cores in the northeast
``main core'' that includes the well-studied B1-b and B1-c cores
\citep{1999sf99.proc..181H,2006ApJ...652.1374M,2010ApJ...712..778H},
to a long filament of starless gas and dust known as the B1 Ridge
\citep{2006ApJ...638..293E}, followed by a collection of gas and dust
clumps in the southwest that includes one protostellar core and
several Class II YSOs. In total, the \textit{Spitzer} c2d team
identified thirteen young stellar objects (YSOs) in our field
\citep{2006ApJ...645.1246J, 2007ApJS..171..447R,2009ApJS..181..321E}.
Twelve dust clumps were identified by \citet{2005A&A...440..151H}
using SCUBA 850~$\mu$m data, while \citet{2006ApJ...646.1009K}
identified five; the difference was attributed to different CLUMPFIND
\citep{1994ApJ...428..693W} thresholds. \citet{2007ApJ...656..293J}
associated an embedded YSO with five of the SCUBA dust clumps, while
the rest of the dust clumps appeared starless.

This diversity of star formation activity along a single filament
makes Barnard~1 a compelling region to spectrally image at high
angular resolution. CLASSy adds to the large collection of work
already done on Barnard~1, and provides the first large area (150
square arcminute), high angular resolution (7\arcsec; 0.008~pc at
235~pc) spectral view of the dense gas ($n$$>$10$^{4-6}$~cm$^{-3}$)
across the entire field. This will allow us to quantify the structure
and kinematics of the gas that is actively participating in the
current generation of star formation in greater detail than ever
before.
    
The primary goals of this paper are to present: (1) the details of our
CLASSy observations and data calibration, (2) an overview of the dense
gas morphology and kinematics in Barnard~1, and (3) a dendrogram
analysis of the \NtwoH{} emission.  In section~2, we describe the
CARMA observations, data calibration, and imaging. The continuum
detections are summarized in section~3. Section~4 provides a
large-scale view of the dense gas structure using integrated intensity
maps. In section~5, we discuss techniques for spectral line fitting
and present centroid velocity and velocity dispersion maps. In
section~6, we create a dendrogram representation of the \NtwoH{} gas
and calculate properties of identified dendrogram structures.  In
section~7, we use the spatial and kinematic properties of
dendrogram-identified structures to construct size-linewidth
relations; these relations are then used to probe the physical and
turbulent nature of Barnard~1.  Section~8 has a summary of the initial
results with a look ahead to upcoming science papers that utilize the
data presented here. Appendix~A compares clump-finding and dendrogram
methods for identifying objects in this region, and Appendix B details
our new dendrogram algorithm that can produce non-binary hierarchies.

\section{OBSERVATIONS}
  
CARMA is an ideal facility for producing molecular line maps that are
sensitive to a wide range of spatial scales. The newly commissioned
CARMA23 mode uses the cross-correlations from all twenty-three
antennas, which increases CARMA's imaging capability from the standard
CARMA15 mode. Using CARMA23 in concert with CARMA's single-dish
capability and fast mosaicing can produce large maps that are
sensitive to line-emission at all spatial scales. We detail the
observations, data calibration, and map making in the sections below.

\subsection{CARMA Interferometric Observations}

We mosaiced an approximately 8\arcmin{} $\times$ 20\arcmin{} area of
Barnard~1 using CARMA. The total observing time (150 hours) was split
evenly between DZ and EZ configurations, which have projected
baselines from about 1-40~k$\lambda$ and 1-30~k$\lambda$,
respectively. DZ configuration observations occurred in the Spring and
Fall of 2012, and EZ configuration observations occurred in the Summer
of 2012.  Table~\ref{tbl:obssum} provides a summary of the
observations and calibrators. The ``Z'' in the DZ and EZ
configurations refers to use of the 23-element observing mode
(CARMA23), which utilizes cross-correlations from all 23 antennas: six
10.4-meter antennas, nine 6.1-meter antennas, and eight 3.5-m
antennas. (Standard CARMA15 observing only includes the 10.4-meter and
6.1-meter antennas.)

\begin{deluxetable}{l c c c c c }
\tabletypesize{\footnotesize}
\tablecaption{Observation Summary}
\tablewidth{0pt}
\setlength{\tabcolsep}{0.03in}
\tablehead
{
 \colhead{Array} & \colhead{Dates} & \colhead{Total Hours} & \colhead{Flux Cal.} & \colhead{Gain Cal.} & \colhead{Mean Flux (Jy)}
}
\startdata    
DZ  & April - May 2012  & 50  & Uranus & 3C84/3C111 & 18.3/3.8 \\
   & October 2012  & 25  & Uranus & 3C84/3C111 & 17.5/2.9 \\
EZ  & July - September 2012  & 75  & Uranus & 3C84/3C111 & 18.9/3.6 \\
\enddata
\vspace{-0.5cm}
\label{tbl:obssum}
\end{deluxetable}

CARMA23 provides more baselines compared with the CARMA15 mode--this
enhances imaging capabilities by filling in more of the
\textit{uv}-plane. It also offers shorter baselines, which are
important for two reasons: (1) more extended emission, if present, is
recovered, and (2) the increased \textit{uv}-coverage improves the
link between the zero spacing information from single-dish and the
interferometer visibilities when doing joint deconvolution.

\begin{figure}[H]
\centering 
\includegraphics[scale=0.5]{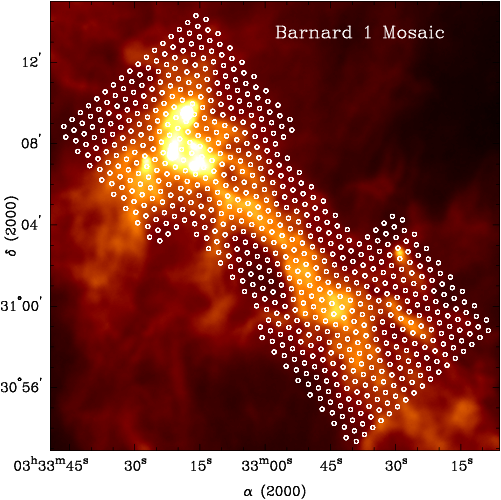}
\caption{\small Mosaic pointing centers overlaid on a \textit{Herschel}
  250~$\mu$m image of Barnard~1 \citep{2010AA...518L.102A}. We sampled
  the field at 31\arcsec{} spacing; the smallest CARMA primary beam,
  from the 10.4-m antennas, is about 77\arcsec{} near 90~GHz. We planned
  the observations to cover the brightest dust emission in the
  Barnard~1 main core, and then cover the filamentary emission that
  extends to the southwest. Additional pointings around the edge aid in
  joint deconvolution by limiting strong emission at the edges of the
  single-dish maps.  }
\label{fig:mosaicpt}
\end{figure}

Our CARMA23 mosaic of Barnard~1 is made up of three adjoining
rectangular regions, each at a position angle of 40\deg{} east of north;
it contains 743 individual pointings in a hexagonal grid with
31\arcsec{} spacing. See Figure~\ref{fig:mosaicpt} for a depiction of
the pointing centers overlaid on a \textit{Herschel} 250~$\mu$m image
\citep{2010AA...518L.102A}. The reference position of the map, which
is the center of the northern rectangle, is at
$\alpha$=03$^{\textrm{\footnotesize h}}$33$^{\textrm{\footnotesize
    m}}$20$^{\textrm{\footnotesize s}}$,
$\delta$=31\deg08\arcmin45\arcsec (J2000); it encompasses the
Barnard~1 main core with the B1-b and B1-c continuum cores. The other
two rectangles extending toward the southwest were chosen to follow
the \textit{Herschel} dust emission.

During each pass through the map, we integrated for 15~seconds on each
mosaic position before moving to the next mosaic position. We used a
newly commissioned ``continuous integration'' technique that improves
on-source efficiency (on-source integration time compared to wall
clock time) of large CARMA mosaics from about 35\% to 48\%. The
improvement comes from continuously taking data, even while antennas
slew to a new mosaic position; the data for a given antenna is
automatically blanked while it is slewing. This technique removes the
built-in software delay associated with a slew to a new target or
mosaic position. The gain in on-source efficiency is critically
important for large mosaic observations, with many moves and short
integration times between moves. This method should not be confused
with on-the-fly mosaicing, where the antennas are continuously moving
while taking data.

In CARMA23 mode, the correlator has four spectral bands in the upper
side band. The lower side band is not present on the 3.5-m telescopes;
it is present but not used on the 10.4 and 6.1-m telescopes.  We tuned
to the $J=1\rightarrow0$ transitions of \NtwoH, \HCO, and HCN, and
used a 500 MHz wide band for calibration and continuum detections.
Table~\ref{tbl:corrsum} has an overview of the correlator setup.  The
molecular lines were each placed in 8~MHz bands, which have 159
channels that are 0.049~MHz wide in 3-bit mode; this corresponds to
$\sim$0.16~\kms{} velocity resolution near 90~GHz.  All hyperfine
components of \NtwoH{} and HCN fit within the 8~MHz bands. However,
there is high-velocity \HCO{} and HCN emission from outflows that lies
outside our bandwidth.

\begin{deluxetable}{l c c c c c c c}
\tabletypesize{\footnotesize}
\tablecaption{Correlator Setup Summary}
\tablewidth{0pt}
\setlength{\tabcolsep}{0.03in}
\tablehead
{
 \colhead{Line} & \colhead{Rest Freq.} & \colhead{No. Chan.} & \colhead{Chan. Width} & \colhead{Vel. Coverage} & \colhead{Vel. Resolution} & \colhead{Chan. RMS} & \colhead{Synth. Beam$^{a}$} \\
 \colhead{} & \colhead{(GHz)} & \colhead{} & \colhead{(MHz)} & \colhead{(\kms)} & \colhead{(\kms)} & \colhead{(\Jybm)} & \colhead{} 

}
\startdata    
N$_{2}$H$^{+}$ & 93.173704 & 159 & 0.049 & 24.82 & 0.157 & 0.14 & 7.6\arcsec{}~$\times$~6.5\arcsec{} \\
Continuum      & 92.7947   &  47 & 10.4 & 1547 & 33.6 & 0.0013 & 8.0\arcsec{}~$\times$~6.2\arcsec \\
HCO$^{+}$      & 89.188518 & 159 & 0.049 & 25.92 & 0.164 & 0.12 & 7.8\arcsec{}~$\times$~6.8\arcsec{}\\
HCN            & 88.631847 & 159 & 0.049 & 26.10 & 0.165 & 0.12 & 7.9\arcsec{}~$\times$~6.8\arcsec{}\\
\enddata
\vspace{-0.5cm} \tablecomments{ $^{a}$In principle, the synthesized
  beam is slightly different for each pointing. Miriad calculates a
  synthesized beam for the full mosaic based on all of the pointings.
}
\label{tbl:corrsum}
\end{deluxetable}

Two team members independently inspected, flagged, and calibrated each
observing track during the 150 hour campaign using MIRIAD
\citep[Multichannel Image Reconstruction, Image Analysis and
  Display;][]{1995ASPC...77..433S}. The flags for each track were then
copied to our MIS \citep[MIRIAD Interferometric and
  Single-Dish;][]{2013AAS...22124009T} pipeline that has the
capability to calibrate all of the tracks with minimal user input and
create combined visibility datasets and maps of all the observed
sources.  Standard interferometric calibration steps (e.g., passband,
gain, flux calibration) were performed in the MIS pipeline calibration
code. We observed a nearby quasar every 16 minutes for gain
calibration; 3C84 was the main gain calibrator, and 3C111 was used
when 3C84 rose above 80\deg{} elevation.  3C84 doubled as a passband
calibrator.  We observed Uranus for absolute flux calibration during
each track and used the {\tt bootflux} task to determine the absolute
flux of the gain calibrators. The flux of 3C84 varied between 16.0 and
19.4~Jy over the course of the campaign, while 3C111 varied between
2.8 and 4.5~Jy. The uncertainty in absolute flux calibration in the
24 combined datasets is about 10\%; hereafter, we only report
statistical uncertainties in quoting errors in measured values.

\subsection{CARMA Single-Dish Observations}

Simultaneously with the interferometric observations, we obtained
CARMA total power observations to recover the line emission resolved
out by the interferometer (i.e., CARMA in standard interferometric
observing mode). CARMA's single-dish mode utilizes the autocorrelation
capability of the correlator and intersperses observations of an
emission-free region between on-source integrations. A total power
spectrum can then be constructed from knowledge of the system
temperature (T$_{\textrm{sys}}$), the emission region (the ON
position), and the emission-free region (the OFF position).

The OFF position for Barnard~1 was in a gap of $^{12}$CO and $^{13}$CO
emission 28\arcmin{} west and 1.5\arcmin{} south of the Barnard~1
mosaic reference position.  A hole in lower density CO gas ensured
that there was no significant dense gas in that region. We integrated
on the OFF position for 30 seconds every 3.5 minutes if the
atmospheric opacity was stable on the 3.5 minute timescale; otherwise,
we observed purely in interferometric mode. Fourteen of 24 tracks with
passing weather grades were observed in single-dish mode. We flagged
the autocorrelation data separately from the cross-correlation data to
account for cases where the opacity deteriorated in a track that was
started in single-dish mode.

We calibrated the autocorrelation data from each antenna in MIRIAD
with two steps: (1) {\tt sinbad} calculated $\mathrm{T_{sys} * (ON - OFF) /
(OFF)}$, and (2) {\tt sinpoly} removed first-order polynomial baselines
from the {\tt sinbad} spectra. We averaged OFF scans on both sides of
an ON scan in the {\tt sinbad} calculation to reduce noise introduced
from OFF scans.

We converted the calibrated single-dish spectra from each of the six
10.4-m antennas to six single-dish data cubes using the {\tt varmaps}
routine. We only used the 10.4-m antennas because they have the
highest angular resolution and hence best overlap with the 3.5-m
baselines in the \textit{uv}-plane. The 10.4-m antennas have a
halfpower beamwidth of 77.3\arcsec{} (\NtwoH), 80.7\arcsec{} (\HCO),
and 81.2\arcsec{} (HCN). The routine gridded each data cube onto
10\arcsec{}~$\times$~10\arcsec{} cells, and used a 50\arcsec{}
smoothing beam to calculate the emission in each cell according to the
emission at each mosaic pointing. The size of the smoothing beam was
determined empirically; a smaller beam increased the final noise in
the maps, while a larger beam smoothed out structure seen in other
single-dish maps of this region. The final halfpower beamwidth of the
calibrated single-dish cubes is the quadrature sum of the original
halfpower beamwidth and the smoothing beam: 92.1\arcsec{} (\NtwoH),
94.9\arcsec{} (\HCO), and 95.4\arcsec{} (HCN).

We scaled the data cubes from all six antennas to a single reference
antenna to account for systematic differences of $\sim$10\% arising
from physical differences in each antenna and from differences in
bandpass shape of each antenna response. We calculated the mean of the
six data cubes using {\tt imstack} to improve the signal-to-noise
ratio and limit antenna-based artifacts. The antenna temperature rms
values in the final cubes are 0.025, 0.027, and 0.026~K for \NtwoH,
\HCO, and HCN, respectively.

\subsection{Joint Deconvolution of Interferometric and Single-Dish Cubes}

The final data product for each observed molecular line transition is
a spectral line cube produced from a joint deconvolution of the
interferometric and single-dish data. The joint deconvolution was done
in MIRIAD as summarized below.

We created the interferometric dirty cube and dirty beam from the
calibrated visibility dataset using {\tt invert} with system
temperature and antenna gain weighting, and Briggs' robustness
parameter \citep{1999ASPC..180..127B} of $-$0.5. We de-selected
10.4-m-and-3.5-m baselines due to the illumination of the first
negative sidelobe of the 10.4-m beam by the 3.5-m beam. The dirty cube
was then cleaned with {\tt mossdi}, a Steer CLEAN algorithm, and the
clean components were carried over to the joint deconvolution to aid
in the convergence to a solution. The single-dish cube was regridded
to the interferometric cube axes, and converted to \Jybm{} units with
a 65~\Jybm~K$^{-1}$ scaling
factor\footnote{$\textrm{http://www.mmarray.org/memos/carma\_memo52.pdf}$}.

The joint deconvolution was done with a maximum entropy algorithm in
MIRIAD, {\tt mosmem}, that used the interferometric dirty cube,
single-dish cube, interferometric dirty beam, single-dish beam, and
interferometric clean components to solve for the maximum entropy
model components.  The final cubes are the dirty cube, minus the
maximum entropy model components convolved by the dirty beam, plus the
maximum entropy model components convolved by the synthesized beam.
The noise levels and synthesize beams for the final data cubes are
given in Table~\ref{tbl:corrsum}.

\subsection{Continuum Mapping}

A 3~mm continuum map was created from the interferometric data in the
500 MHz window. We created and cleaned the dirty map in MIRIAD using
{\tt invert} and {\tt mossdi}, and restored with a synthesized beam of
8.0\arcsec{}~$\times$~6.2\arcsec.  The rms in the calibrated continuum
map is $\sim$1.3 m\Jybm{}.

The autocorrelation data from the 500~MHz band was not used in
continuum mapping because single-dish continuum observations require
fast chopping between on-position and off-position to cancel out
variable sky emission; this observing technique is not available at CARMA.

\section{CONTINUUM RESULTS}

\begin{figure}[H]
\centering \includegraphics[scale=0.8]{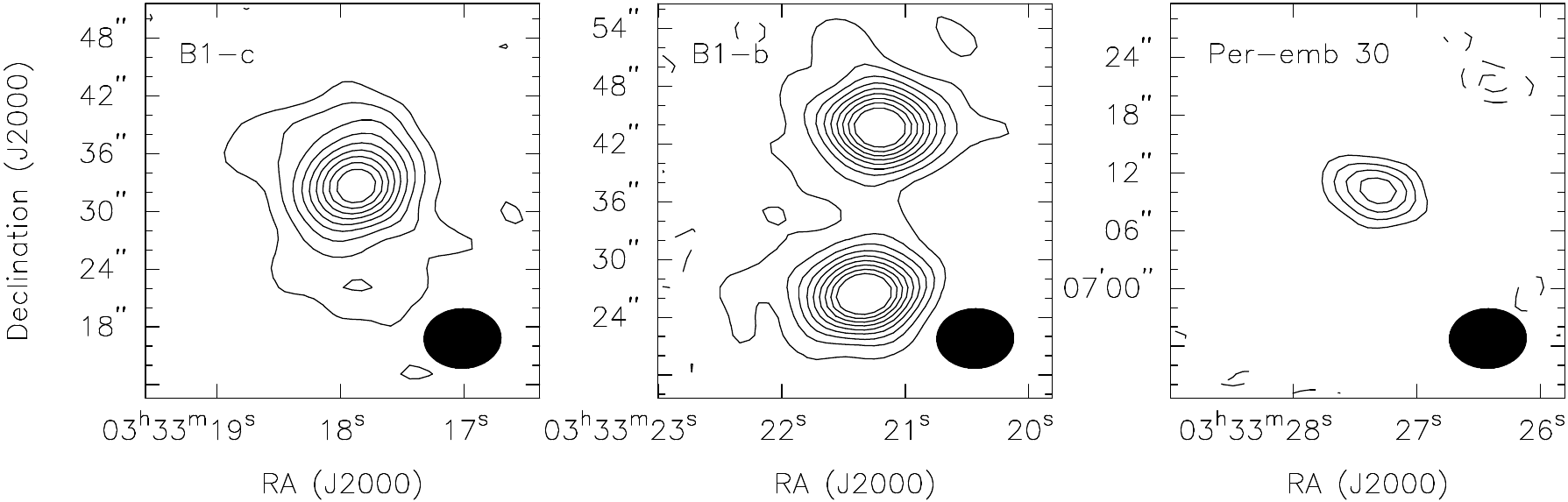}
\caption{\small The four continuum detections above 5-$\sigma$ in our
  field. The synthesized beam is 8.0\arcsec{}~$\times$~6.2\arcsec, and
  the 1-$\sigma$ sensitivity is $\sim$1.3~m\Jybm. B1-c and B1-b were
  previously found to have compact continuum emission associated
  with young stars. The contour levels in those maps are
  ($\pm$)2, 4, 6, 8, 10, 12, 14, 16, 18 times 1-$\sigma$. The compact
  emission within Per-emb 30 is a new detection that peaks at
  5.8-$\sigma$. Contour levels in that map are ($\pm$)2, 3, 4, 5 times
  1-$\sigma$. The negative contours are represented by dashed lines.}
\label{fig:cont}
\end{figure}

We detected four compact continuum sources above the 5-$\sigma$ level;
all are associated with young protostars in the main
core. Figure~\ref{fig:cont} shows 3~mm continuum images toward the
previously known Class 0 source, B1-c \citep{2006ApJ...652.1374M}, and
the Class 0 double source, B1-b \citep{1999sf99.proc..181H}, along
with a new 5.8-$\sigma$ detection toward a Class I source, SSTc2d
J033327.3+310710 \citep[also at the position of the lower-resolution
  Per-emb 30 dust source in][]{2009ApJ...692..973E}.

The structure around B1-c and B1-b is clearly resolved; Per-emb~30
appears to be nearly unresolved. We used MIRIAD's {\tt imfit} to
determine the position, peak brightness, total flux density, size,
and position angle of each source with an elliptical Gaussian
approximation (see Table~\ref{tbl:cont}). The measured position errors
are 0.2-0.7\arcsec. The positions of B1-c and B1-b agree with other
interferometric observations of these sources
\citep{2006ApJ...652.1374M, 2013ApJ...768..110C, 2013ApJ...766..131H}.

\begin{deluxetable}{l c c c c c c c c c }
\tabletypesize{\tiny} \tablecaption{Observed Properties of Continuum Detections} \tablewidth{0pt} \setlength{\tabcolsep}{0.03in}
\tablehead{ 
\colhead{Source} & \colhead{{Position}} & \colhead{{Pk. Bright.}} & \colhead{{Total S$_{\nu}$}} & \colhead{{Ang. Size}} & \colhead{{P.A.}} & \colhead{{Decon. Size}} & \colhead{{Decon. P.A.}} & \colhead{{Lin. Size}} & \colhead{{Mass}} \\
\colhead{Name} & \colhead{{(h:m:s, d:\arcmin:\arcsec)}} & \colhead{{(m\Jybm)}} & \colhead{{(mJy)}} & \colhead{{(\arcsec)}} & \colhead{{(\deg)}} & \colhead{{(\arcsec)}} & \colhead{{(\deg)}} & \colhead{{(AU)}} & \colhead{{(\Msun)}}\\
\colhead{(1)} & \colhead{{(2)}} & \colhead{{(3)}} & \colhead{{(4)}} & \colhead{{(5)}} & \colhead{{(6)}} & \colhead{{(7)}} & \colhead{{(8)}} & \colhead{{(9)}} & \colhead{{(10)}}
}
\startdata 
B1-c & 03:33:17.91, +31:09:32.8       & 23.0 $\pm$ 1.8 & 52.6 $\pm$ 4.2 & 10.7~$\times$~10.7 & -33 & 8.7~$\times$~7.1 & -1   & 2040~$\times$~1670 & 1.04~$\pm$~0.08 \\
B1-b South & 03:33:21.37, +31:07:26.4 & 29.1 $\pm$ 2.3 & 42.1 $\pm$ 3.3 & 9.8~$\times$~7.4 & -79 & 5.9~$\times$~3.6  & -65 & 1390~$\times$~850  & 0.84~$\pm$~0.07 \\
B1-b North & 03:33:21.23, +31:07:43.7 & 27.5 $\pm$ 2.2 & 43.0 $\pm$ 3.3 & 10.1~$\times$~7.7 & 89 & 6.2~$\times$~4.5  & 87  & 1460~$\times$~1060 & 0.85$\pm$~0.07 \\
Per-emb~30 & 03:33:27.35, +31:07:10.0 & 6.9 $\pm$ 0.8  & 8.9 $\pm$ 1.0  & 9.7~$\times$~6.6 & 79  & 5.5~$\times$~1.7  & 69  &  1290~$\times$~400 & 0.18~$\pm$~0.02 \\
\tableline
\noalign{\smallskip}
W1 & 03:33:14.8, +31:07:13 & 4.1 $\pm$ 1.3 & ... & ... & ... & ... & ... &  ...  & $^{a}$0.08~$\pm$~0.03 \\ 
W2 & 03:33:16.7, +31:06:53 & 4.4 $\pm$ 1.3 & ... & ... & ... & ... & ... &  ...  & $^{a}$0.09~$\pm$~0.03
\enddata
\vspace{-0.5cm} \tiny \tablecomments{ (3) Peak brightness, (4)
  Total flux density, (5) Major and minor axes (FHWM), (6) Position
  angle (east of north), (7) Deconvolved major and minor axes
  (FHWM), (8) Deconvolved position angle, (9) Linear size computed
  from deconvolved size assuming a distance of 235~pc, (10) Mass
  calculated using assumptions in the text ($^{a}$lower-limit mass for
  weak detections that uses the peak brightness instead of the
  total flux density).}
\label{tbl:cont}
\end{deluxetable}

With linear sizes of 1000-2000~AU, the 3~mm emission is arising from
compact cores associated with forming stars. In the optically thin
limit, under the assumption of a single temperature, core mass is
related to the total flux density as
\begin{equation}
 M = \frac{F_{\nu}D^{2}}{\kappa_{\nu} B_{\nu}(T_{\mathrm{d}})},
\end{equation}
where $M$, $F_{\nu}$, $D$, $\kappa_{\nu}$, and
$B_{\nu}(T_{\mathrm{d}})$, are respectively the total mass, total
observed flux density, distance, mass opacity including dust and gas
(assuming a gas-to-dust ratio of 100), and blackbody intensity at dust
temperature, $T_{\mathrm{d}}$.  We assume $T_{\mathrm{d}}$ = 20~K. To
estimate $\kappa_{\nu}$, we assumed a power law opacity curve,
$\kappa_{\nu} = \kappa_{\mathrm{o}}(\nu/\nu_{\mathrm{o}})^{\beta}$,
where $\nu_{\mathrm{o}}$=1000~GHz and
$\kappa_{\mathrm{o}}$=0.1~cm$^{2}$~g$^{-1}$
\citep{1990AJ.....99..924B}. For a $\beta$ of 1.5, $\kappa_{\nu}$ is
0.0028~cm$^{2}$~g$^{-1}$ at 92.79~GHz. The core masses under these
assumptions are listed in Table~\ref{tbl:cont}; statistical errors are
reported using the uncertainty in the total flux density. The B1-c
mass estimate is about three times lower than the mass reported in
\citet{2006ApJ...652.1374M}, primarily due to differences in the
observed total flux density near 3.3~mm. The B1-b mass estimates are
several times larger than results from \citet{2013ApJ...766..131H},
who observed the sources at 1~mm with higher angular resolution; the
disparity increases when adopting their assumed $\beta$ and
$T_{\mathrm{d}}$ values. Since they derived deconvolved sizes of
$\sim$300-500~AU, it is most likely that their mass estimates are not
including extended emission in the protostellar envelope.

In addition to these relatively strong detections, we detected two
continuum peaks greater than 3-$\sigma$ (labeled W1-2 in
Table~\ref{tbl:cont}) that are coincident with a source in at least
one \textit{Spitzer} or \textit{Herschel} band. The positions, peak
brightnesses, and lower-limit masses for these sources are listed in
Table~\ref{tbl:cont}.  We determined position using {\tt imfit} with
an elliptical Gaussian approximation. Peak brightness was defined as
the maximum pixel value within the 3-$\sigma$ contour of the source,
with error equal to the 1-$\sigma$ sensitivity of the continuum
map. The lower-limit mass was calculated from the peak brightness. W1
is coincident with SSTc2d J033314.3+310710 (Per-emb~6), and W2 is
coincident with SSTc2d J033316.4+310652 (Per-emb 10).  Deeper,
follow-up observations are needed to confirm these detections and
calculate physical properties of the sources.

\section{MORPHOLOGY OF DENSE MOLECULAR GAS}

\begin{figure}[]
\centering \includegraphics[scale=0.85]{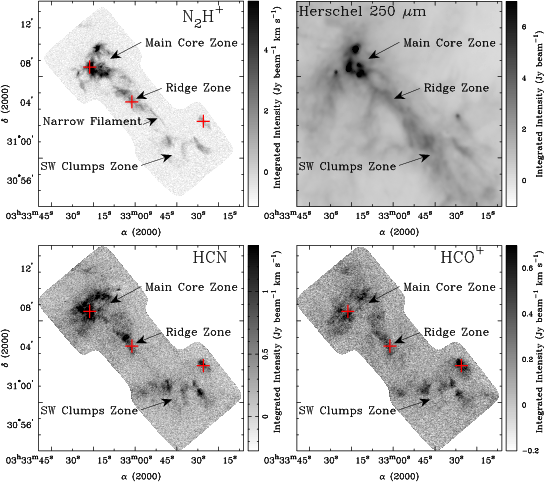}
\caption{ \small Integrated intensity maps of \NtwoH, HCN, and \HCO{}
  ($J=1\rightarrow0$) emission towards Barnard~1, with a
  \textit{Herschel} 250~$\mu$m map.  Channels containing outflow
  emission are excluded from these maps, which is why the narrow
  filament (see Section~4.2) appears only in \NtwoH.  The rms values
  of the \NtwoH, HCN, and \HCO{} maps are 0.17, 0.10, and
  0.06~\Jybmkms, respectively. We break the region into three zones
  based on qualitative features in the dense gas maps. Crosses in the
  maps represent the locations of spectra shown in
  Figure~\ref{fig:spectra}. FITS cubes of the PPV data used to make
  this figure are available in the online journal.}
\label{fig:birdseye}
\end{figure}

\begin{figure}[]
\centering 
\includegraphics[scale=0.85]{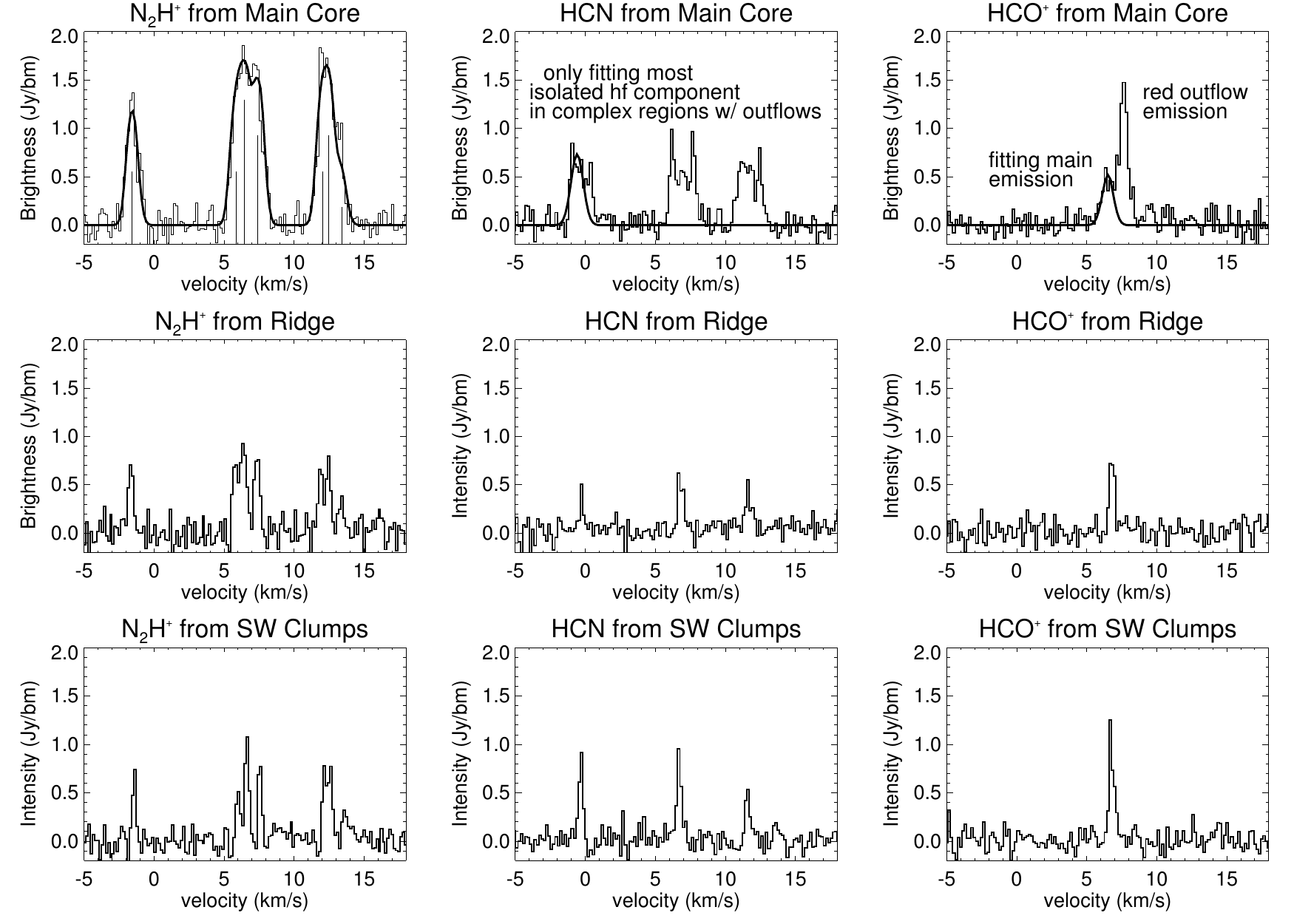}
\caption{ \small Example spectra for all three molecules from each
  zone. The spectra are averaged over one synthesized beam, and the
  positions are: $\alpha$ =
  03$^{\textrm{h}}$33$^{\textrm{m}}$21.633$^{\textrm{s}}$, $\delta$ =
  31\deg07\arcmin38.06\arcsec{} for the Main Core Zone, $\alpha$ =
  03$^{\textrm{h}}$33$^{\textrm{m}}$01.482$^{\textrm{s}}$, $\delta$ =
  31\deg04\arcmin05.09\arcsec{} for the Ridge Zone, and $\alpha$ =
  03$^{\textrm{h}}$32$^{\textrm{m}}$27.312$^{\textrm{s}}$, $\delta$ =
  31\deg02\arcmin05.36\arcsec{} for the SW Clumps Zone.  We fit the
  spectra and present maps of centroid velocity and velocity
  dispersion in section~5.  The fits to the main core zone spectra are
  shown with solid black lines.  The \NtwoH{} has seven resolvable
  hyperfine components in narrow-line regions, and HCN has three
  resolvable hyperfine components.  The hyperfine structure of
  \NtwoH{} is shown with vertical bars set within the main core zone
  spectrum---the relative brightnesses are representative of the
  hyperfine structure, but the absolute values were picked for ease of
  visualization.  As described in section~5, we exclude outflow
  emission from HCN and \HCO{} when doing these fits.}
\label{fig:spectra}
\end{figure}

Figure~\ref{fig:birdseye} shows our \NtwoH, HCN, and \HCO
($J=1\rightarrow0$) integrated intensity maps, along with a
\textit{Herschel} 250~$\mu$m view of Barnard~1 for a qualitative
gas-dust comparison. The angular resolution of the dust map is
18.1\arcsec{} in comparison to our $\sim$7\arcsec{} resolution.  To
facilitate the discussion later in this section, we identify three
zones of emission: the main core zone, the ridge zone, and the SW
clumps zone. Figure~\ref{fig:spectra} shows example spectra from each
zone at the locations marked by crosses in Figure~\ref{fig:birdseye},
and the next three sections discuss the three zones in detail.

The \NtwoH{} map was integrated over all seven hyperfine components
over velocity ranges from 14.044 to 10.745~\kms, 8.860 to 4.775~\kms,
and $-$0.951 to $-$2.766~\kms. The HCN map was integrated over all
three hyperfine components, excluding channels with outflows, from
12.117 to 10.961~\kms, 7.326 to 6.005~\kms, and 0.057 to
$-$1.264~\kms. The \HCO{} map was integrated from 7.161 to 5.849~\kms,
again excluding outflow channels.  We integrated \NtwoH{} over a wider
range of velocities compared to the other molecules to include a
narrow, redshifted filament, which can be seen along the southeastern
edge of the thicker \NtwoH{} filament in the ridge zone. This
redshifted filament, discussed later in this section, is also detected
in HCN and \HCO, but overlaps in velocity space with HCN and \HCO{}
outflow channels that were excluded from these integrated intensity
maps. The rms of the \NtwoH, HCN, and \HCO{} integrated intensity maps
in Figure~\ref{fig:birdseye} are 0.17, 0.10, and 0.06~\Jybmkms,
respectively.

The peak brightness temperature for a single channel (including
channels with outflows) in the entire field occurs towards the B1-c
continuum core; it is 9.5~K, 9.7~K, and 7.1~K in our synthesized beam
for \NtwoH{} (2.85~\Jybm{} to K conversion factor), HCN (2.91~\Jybm{}
to K conversion factor), and \HCO{} (2.91~\Jybm{} to K conversion
factor), respectively.

\subsection{Main Core Zone}

\begin{figure}[H]
\centering 
\includegraphics[scale=0.75]{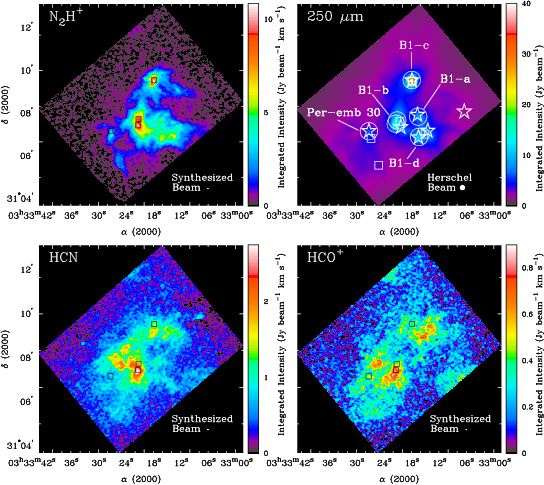}
\caption{ \small Integrated intensity maps of \NtwoH, HCN, and \HCO{}
  ($J=1\rightarrow0$) emission toward the main core zone, along with
  \textit{Herschel} 250~$\mu$m.  Channels containing outflow emission
  are excluded from the integrated intensity maps. The angular
  resolution of each map is marked with a beam in the lower right
  corner---the CARMA synthesized beam is $\sim$7\arcsec{}. The
  \textit{Herschel} map shows the location of \textit{Spitzer} YSOs
  (stars), Bolocam 1.1~mm cores (squares), and SCUBA 850~$\mu$m cores
  (circles); all YSOs in this region are Class 0/I. Major dust clumps
  mentioned in the text are labeled in the \textit{Herschel} map. The
  black boxes in the integrated intensity maps show the locations of
  the compact continuum sources that were discussed in section~3.  The
  rms of the \NtwoH, HCN, and \HCO{} integrated intensity maps are
  0.17, 0.10, and 0.06~\Jybmkms, respectively.}
\label{fig:mom0maincore}
\end{figure}

\begin{figure}[H]
\centering 
\includegraphics[scale=0.75]{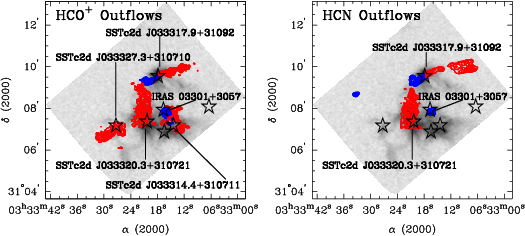}
\caption{\small \HCO{} and HCN outflows overlaid on \NtwoH{}
  integrated intensity emission. Each outflow was integrated over its
  own range of velocities to minimize noise introduced from
  emission-free channels. We are plotting contours of 0.3, 0.4, 0.5,
  0.6, 0.7, 0.8, 0.9 times the peak integrated intensity of the
  outflow, except for the HCN bipolar outflow from SSTc2d
  J033317.9+31092 where we extend down to 0.1 and 0.2 times the
  peak. See Table~\ref{tbl:outflow} for the velocity ranges and peak
  integrated intensity of each outflow---the outflows are identified
  in the table by the nearest infrared source plotted in this
  figure. \textit{Spitzer} YSO positions are indicated with black
  stars; all YSOs are Class 0/I in this region.}
\label{fig:outflows}
\end{figure}

The main core zone, shown in Figure~\ref{fig:mom0maincore}, is the
area of strongest dust emission seen by \textit{Herschel}; it contains
the largest cluster of young protostellar objects, and is the only
region where we have detected compact continuum sources (see
section~3) and outflows (see Figure~\ref{fig:outflows}). The \NtwoH{}
emission in this zone follows the overall structure of the dust,
although our improved angular resolution reveals more detail than ever
before. The HCN and \HCO{} emission not associated with outflows is
much weaker compared to \NtwoH, and it does not follow the
morphological structure of the \NtwoH{} or the dust.
Figure~\ref{fig:mom0maincore} shows the locations of \textit{Spitzer}
YSO candidates \citep{2006ApJ...645.1246J}, Bolocam 1~mm clumps
\citep{2006ApJ...638..293E}, and SCUBA 850~$\mu$m clumps
\citep{2005A&A...440..151H}.

The spectra from this zone are the most complex in the Barnard~1 field
-- Figure~\ref{fig:spectra} shows a sample \NtwoH{} spectrum from near
the B1-b core that exhibits broad lines, while the HCN and \HCO{}
spectra from the same location show evidence of red-wing outflow
emission overlapping with B1-b. The \NtwoH{} emission towards B1-b
shows two resolved peaks. There are emission peaks in HCN and \HCO{}
towards B1-b South, but not B1-b North (see
Figure~\ref{fig:mom0maincore}). This agrees with findings from
\citet{2013ApJ...766..131H} and supports the idea that carbon-bearing
species are depleted around B1-b North, suggesting it is the younger
source of the two.

The \NtwoH{} gas emission is very strong at the location of B1-c, while
\HCO{} is only detected in outflows; this agrees with observations from
\citet{2006ApJ...652.1374M}. HCN gas does exist at the location of
B1-c, but is not a strong peak relative to the rest of the main core,
like we see with \NtwoH{}. HCN is also detected in outflow emission.

There are seven \textit{Spitzer} Class 0/I YSOs in this zone. Two are
known to be associated with the B1-c and Per-emb~30 compact continuum
cores, which are sites of strong \NtwoH{} emission. There are four
nearby the larger scale B1-a, B1-b, and B1-d dust clumps, which are
also areas of strong \NtwoH emission. The seventh YSO, which lies west
of the strongest dense gas in this zone, only has weak dust and gas
emission associated with it. Several of these YSOs are driving
outflows, which are discussed later in this section, and identified in
Figure~\ref{fig:outflows} and Table~\ref{tbl:outflow}.

There are five Bolocam 1.1~mm cores and six SCUBA 850~$\mu$m cores in
this region. All of these lower resolution dust cores are near
\NtwoH{} emission peaks, but not necessarily near HCN or \HCO{}
peaks. There are several peaks of HCN and \HCO{} that are not
associated with strong dust emission.

Figure~\ref{fig:outflows} shows \HCO{} and HCN dense gas outflows,
which are only detected in this zone.  In the \HCO{} map,
SSTc2d~J033317.9+31092 (B1-c) is clearly driving a bipolar outflow. We
detect another bipolar \HCO{} outflow, likely associated with the IRAS
03301+3057 YSO based on a similar detection in CO by
\citet{1997ApJ...478..631H}. There is a red outflow just to the west
of SSTc2d J033314.4+310711, another to the north SSTc2d
J033320.3+310721, and a third to the southeast of SSTc2d
J033327.3+310710.

In the HCN map, we again see the bipolar outflow from
SSTc2d~J033317.9+31092 (B1-c), although the HCN additionally traces an
extension of the red part of the outflow to the west, and an
associated knot of emission to the east. The western extension flares
out into two separate, high-velocity outflows that extend beyond the
edge of our spectral window. The eastern knot is likely a bright bow
shock from the edge of the outflow interacting with cloud material.
We confirmed the association of these extended features with the main
outflow using \textit{Spitzer} images that trace the shocked gas.  The
outflows from IRAS 03301+3057 and SSTc2d J033320.3+310721 are also
detected in HCN, while those from J033314.4+310711 and
J033327.3+310710 are not.  

\begin{deluxetable}{l c c c}
\tabletypesize{\footnotesize}
\tablecaption{Outflow Identification}
\tablewidth{0pt}
\setlength{\tabcolsep}{0.03in}
\tablehead
{
 \colhead{Source Identifier$^{a}$} & \colhead{Color} & \colhead{Integrated Velocity Range} & \colhead{Peak Integrated Intensity} \\
 \colhead{} & \colhead{(Red or Blue)} & \colhead{(\kms)} & \colhead{(\Jybmkms)}}
\startdata   
\noalign{\smallskip}
\multicolumn{4}{c}{\HCO}\\
\tableline
\noalign{\smallskip} 
SSTc2d J033317.9+310932 (B1-c, Per-emb 29) & Red & 9.62--7.98 & 1.0 \\
SSTc2d J033317.9+310932 (B1-c, Per-emb 29) & Blue & 5.36--3.72 & 1.9 \\
IRAS 03301+3057 (B1-a, Per-emb 40) & Red & 7.66-7.49 & 0.5 \\
IRAS 03301+3057 (B1-a, Per-emb 40) & Blue & 5.36--1.26  & 2.4 \\
SSTc2d J033327.3+310710 (Per-emb 30) & Red & 7.82--7.32 & 0.8 \\ 
SSTc2d J033320.3+310721 (Per-emb 41)  & Red & 9.95--7.33 &  1.8 \\
SSTc2d J033314.4+310711 (Per-emb 6)  & Red & 13.73--7.82 &  4.4 \\
\noalign{\smallskip}
\tableline
\noalign{\smallskip}
\multicolumn{4}{c}{HCN}\\
\tableline
\noalign{\smallskip} 
SSTc2d J033317.9+310932 (B1-c, Per-emb 29)  & Red & 18.89--11.78, 10.13--7.66 & 12.0 \\
SSTc2d J033317.9+310932 (B1-c, Per-emb 29) & Blue & 10.63--7.66, 5.67--1.54, $(-)1.26$--$(-)3.08$ & 10.1 \\
SSTc2d J033317.9+310932 (B1-c, Per-emb 29) & Blue$^{b}$ & $(-)1.26$--$(-)5.72$ & 3.3 \\
IRAS 03301+3057 (B1-a, Per-emb 40) & Blue & 10.30--9.64, 5.34--4.19  & 2.1 \\
SSTc2d J033320.3+310721 (Per-emb 41) & Red & 12.61--12.28, 8.48--7.33, 0.72--0.22 &  1.5
\enddata
\vspace{-0.5cm} \tiny \tablecomments{ (a) Outflows identified by the
  location of the nearest infrared source. Supplemental Per-emb source
  identifiers are from \citet{2009ApJ...692..973E}. \\ (b) This
  emission corresponds to the blue knot seen in the HCN map of
  Figure~\ref{fig:outflows}, east of the main part of the SSTc2d
  J0333317.9+310932 outflow.}
\label{tbl:outflow}
\end{deluxetable}

\clearpage

\subsection{Ridge Zone}

\begin{figure}[]
\centering 
\includegraphics[scale=0.75]{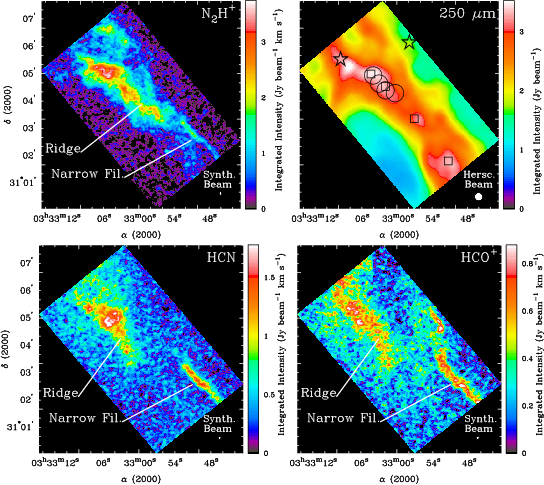}
\caption{\small Integrated intensity maps as in
  Figure~\ref{fig:mom0maincore}, but for the ridge zone. Since the
  ridge zone does not have outflows, we included extra channels that
  were excluded from Figure~\ref{fig:mom0maincore} to capture the
  narrow filament that is redshifted by $\sim$1.5~\kms{} from the rest
  of the Barnard~1 gas. The rms values of these \NtwoH, HCN, and \HCO{}
  integrated intensity maps are 0.17, 0.12, and 0.08~\Jybmkms,
  respectively.}
\label{fig:mom0central}
\end{figure}

The ridge zone, shown in Figure~\ref{fig:mom0central}, lies southwest
of the main core. The most striking feature is the backbone of the B1
Ridge that appears as a structure (``Ridge'' in the figure) that
measures approximately 260\arcsec{} long by 80\arcsec{} wide in the
\NtwoH{} map. This corresponds to about 0.30~pc long by 0.09~pc wide
at a distance of 235~pc. This structure has several Bolocam and SCUBA
cores running along its spine, which are strongly clustered in the
northeast half where gas emission from all molecules is the
strongest. A single Class 0/I YSO sits at the northeast edge of the
structure---none of the dust cores along the structure have associated
protostars, and they are considered pre-stellar. There is a single
Bolocam core at the southwestern end of the structure, where the
\textit{Herschel} map peaks in an area with little molecular emission;
multi-wavelength \textit{Herschel} images show that this region is
brighter at shorter wavelengths compared to the northeastern section
of the structure. This means it could be a region of slightly higher
temperature, which would explain a lower abundance of dense molecular
gas. Figure~\ref{fig:spectra} shows sample spectra from along this
structure.

\begin{figure}[]
\centering 
\includegraphics[scale=0.85]{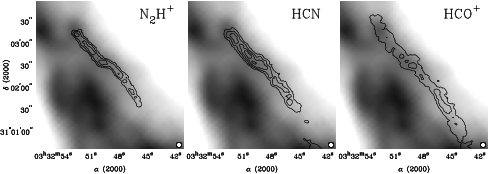}
\caption{\small We discovered a narrow filament that is offset by
  1.5~\kms{} from the rest of the gas in Barnard~1. The three figures
  show the filament detected in \NtwoH, HCN, and \HCO, overlaid on the
  \textit{Herschel} 250~$\mu$m dust image. The contours represent
  integrated intensity emission at 0.3, 0.5, 0.7, and 0.9 times the
  peak integrated intensity of the HCN filament, which is
  1.36~\Jybmkms.}
\label{fig:fil2}
\end{figure}

A second molecular feature in this zone is a newly discovered filament
that runs parallel to the main ridge (``Narrow Fil'' in
Figure~\ref{fig:mom0central}). This filament is extremely narrow and
is offset from the rest of the gas in velocity space by about
1.5~\kms. Figure~\ref{fig:fil2} shows molecular line contours overlaid
on a \textit{Herschel} 250~$\mu$m map. The filament is about
20\arcsec{} wide and 2.5\arcmin{} long, with a small kink half-way
along its length. At a distance of 235~pc, this corresponds to
0.022~pc wide by 0.17~pc long. The peak integrated intensity of the
filament is 1.2~\Jybmkms, 1.4~\Jybmkms, and 0.9~\Jybmkms, for \NtwoH,
HCN, and \HCO, respectively.  It is not possible to identify the
filament in \textit{Herschel} maps because it is extremely narrow and
lies along the same line of sight as the western edge of the B1 Ridge.
We briefly discuss the kinematics and possible formation mechanisms of
the filament in section~5.

The third molecular feature is gas in the northern part of the zone
that provides a tenuous link between the main ridge structure and the
main core zone. A Class 0/I YSO is located off the southwestern edge
of this feature, but there are no dust condensations directly
associated with it.

\subsection{SW Clumps Zone}
\begin{figure}[]
\centering 
\includegraphics[scale=0.75]{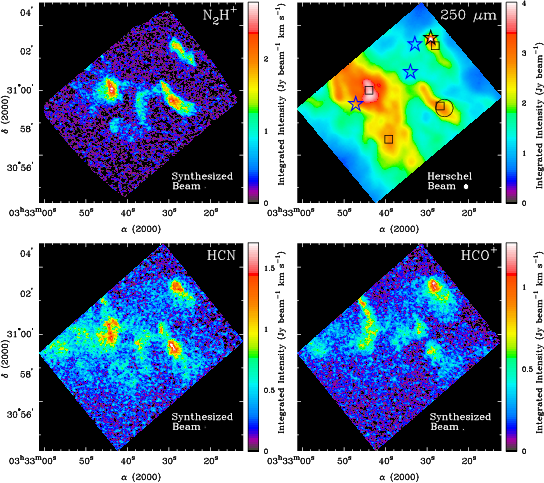}
\caption{\small Same as Figure~\ref{fig:mom0maincore}, but for the SW
  clumps zone. The most western YSO in this zone is a Class 0/I
  source, while the others are Class II sources (colored blue in the
  online version). Like Figure~\ref{fig:mom0central}, since this zone
  does not have outflows, we included extra channels to capture the
  southwestern edge of the narrow filament (best seen in the \HCO{}
  map) that is redshifted by $\sim$1.5~\kms{} from the rest of the
  Barnard~1 gas.  The rms values of these \NtwoH, HCN, and \HCO{}
  integrated intensity maps are 0.17, 0.12, and 0.08~\Jybmkms,
  respectively.}
\label{fig:mom0southern}
\end{figure}

The SW clumps zone lies southwest of the B1 Ridge, and is shown
in Figure~\ref{fig:mom0southern}. There are five distinct clumps of
\NtwoH{} emission, with a sixth clump only detected in HCN and \HCO.
There are three Class II YSOs in this zone that do not correlate with
any gas or dust peaks, which is expected for more evolved
protostars. There is one Class 0/I YSO in this zone that lies at the
northeast tip of the western-most clump.  The strongest integrated
\HCO{} emission across the entire Barnard~1 field is located in this
clump---Figure~\ref{fig:spectra} shows sample spectra from it. The gas
morphology of this zone is very different from the main core and ridge
zones. Those zones have dense gas peaks embedded within
lower-intensity, larger-scale dense gas structures, while the dense
gas peaks in this zone are not joined at weaker emission levels.

\section{KINEMATICS OF DENSE MOLECULAR GAS}
\label{section:kin}
We modeled the spectra in our position-position-velocity (PPV) cubes
to derive the centroid velocity and velocity dispersion for the
emission from each molecule along every line of sight.

The \NtwoH ($J=1\rightarrow0$) line is made up of seven resolvable
hyperfine components.  We model all components simultaneously and
assume they are Gaussians with the same dispersion and excitation
conditions. The line opacity as a function of velocity, $\tau(v)$,
is modeled as:
\begin{equation}
 \tau(v) = \tau\sum\limits_{i=0}^7 C_{i} e^{-(v -
  V_{\mathrm{lsr}} - v_{i})^{2}/2\sigma^{2}},
\end{equation}
where $\tau$ is the total opacity of the emission, $C_{i}$ is the
weighted strength of the $i^{th}$ hyperfine component such that the
sum of component strength is unity, \vlsr{} is the centroid velocity
of the observed emission, $v_{i}$ is the rest velocity of the $i^{th}$
hyperfine component, and $\sigma$ is the velocity dispersion of each
hyperfine component. We assigned $C_{i}$ and $v_{i}$ according to the
CDMS catalog listings on the Splatalogue
webpage\footnote{$\textrm{http://splatalogue.net}$} -- $v_{i}$ was
calculated from the listed $\nu_{i}$ using the rest frequency of our
observations and the radio Doppler formula. A full observed spectrum
is then modeled as
\begin{equation}
 F(v) = \frac{2k\nu^2\Omega(v)}{c^2} \times T_{\mathrm{ex}}(1 -
 e^{-\tau(v)}),
\end{equation}
where $k$ is Boltzmann's constant, $\Omega$ is the synthesized beam
area, $v$ is the velocity of the observations, and $T_{\textrm{ex}}$
is the excitation temperature. 

The four free parameters for the fit are: \vlsr, $\sigma$, $\tau$, and
$T_{\mathrm{ex}}$. We focus on the two kinematic parameters in the
following sections.  We do not address $\tau$ and $T_{\mathrm{ex}}$
because they are semi-degenerate, and breaking the degeneracy comes
from an accurate understanding of the optical depth, which is beyond
the scope of this initial paper.

\begin{figure}[H]
\centering 
\includegraphics[scale=0.85]{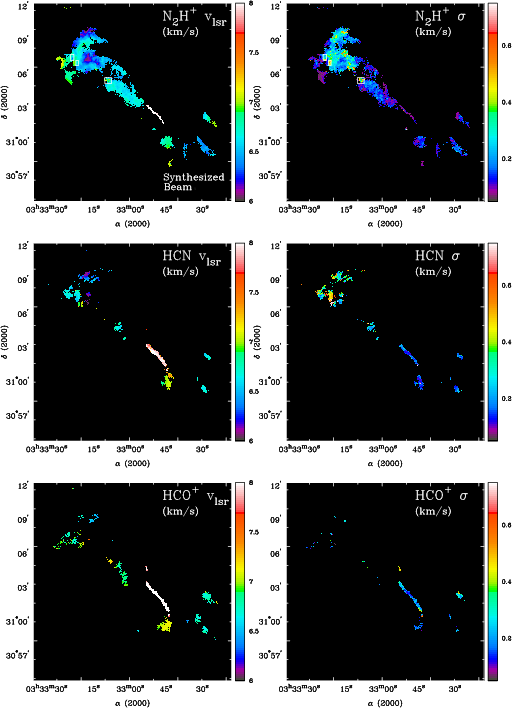}
\caption{\small \textit{Left:} Centroid velocity maps of \NtwoH, HCN,
  and \HCO{} ($J=1\rightarrow0$) emission, from top to
  bottom. \textit{Right:} velocity dispersion maps of \NtwoH, HCN, and
  \HCO{} ($J=1\rightarrow0$) emission, from top to bottom. We masked
  these maps at different levels (see section~5 text) to visualize
  only statistically robust kinematic results. The color scales are
  the same across molecules. The $\sim$7\arcsec{} synthesized beam is
  plotted as a very small white circle in the lower right corner of
  the \NtwoH centroid velocity map. The small, white rectangles
    in the \NtwoH{} maps identify regions that likely contain two velocity components along the-of-sight and are excluded from analysis later in the paper.}
\label{fig:vlsr}
\end{figure}

In the line fitting procedure, a model spectrum is first created from
Equations 2 and 3 with ten times the velocity resolution of our
observed data. We then bin this high-velocity-resolution model
spectrum according to our observations' velocity resolution, and
compare that channelized model spectrum to an observed line. The
fitting is done in IDL with the MPFIT package
\citep{2009ASPC..411..251M}, which performs non-linear least-squares
minimization of the model fit to an observed spectrum---the outputs
are the best-fit values and errors of the four free parameters.
Figure~\ref{fig:vlsr} (top) shows the best-fit centroid velocity
(\vlsr) and velocity dispersion ($\sigma$) maps for \NtwoH. The pixels
that have data values in these maps represent spectra that pass two
robustness criteria: peak signal-to-noise greater than five, and
integrated intensity greater than four times the rms of the \NtwoH{}
integrated intensity map.

The vast majority of the field only contains one resolved
  velocity component along each line-of-sight, and can be adequately
  fit with the procedure described above. However, we manually
  inspected the field and identified four locations with strong
  evidence for having two resolved velocity components. Three of those
  locations are marked with white rectangles in Figure~\ref{fig:vlsr},
  and are excluded from the analysis later in the paper. The fourth
  location is the northeastern part of the narrow filament in the
  ridge zone; we manually inspected this region and removed the small
  number of pixels with confused spectra from the \NtwoH{} maps.

The HCN and \HCO{} emission was fit in a similar way; the HCN line is
made up of three resolvable hyperfine components, and \HCO{} has no
hyperfine structure. Complications arose from outflows in the main
core zone. For lines of sight complicated by outflows, we masked
channels with outflow emission before fitting the spectral line
representing the non-outflow gas. This was only done if the outflow
emission was clearly defined in velocity space---we did not fit lines
where outflow emission blended with emission nearest to the centroid
velocity of the cloud. For HCN lines complicated by outflows, we only
fit the highest frequency hyperfine component as it is more isolated
in frequency/velocity space from the other two hyperfine components.

The middle and bottom rows of Figure~\ref{fig:vlsr} show the best-fit
centroid velocity and velocity dispersion maps for HCN and \HCO{},
respectively.  The integrated intensity criterion is the same as
  above. However, the peak signal-to-noise criterion is eight for HCN,
  and ten for \HCO, to account for increased uncertainty in fit values
  with decreasing number of hyperfine components. It is clear from the
  sparseness of the maps that there are far fewer HCN and \HCO{}
  regions strong enough to get reliable kinematic measurements
  compared to \NtwoH{}. The \HCO{} and HCN data also show evidence for
  two velocity components in the same regions as \NtwoH{}.

\begin{figure}[H]
\centering 
\includegraphics[scale=0.75]{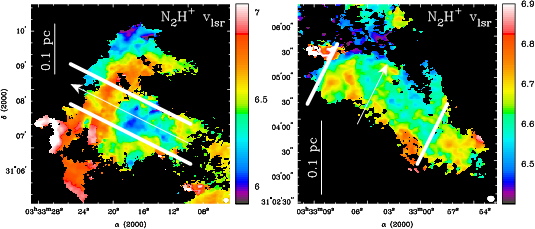}
\caption{\small \textit{Left:} Zoom-in of the main core zone centroid
  velocities presented in Figure~\ref{fig:vlsr}, showing the complex
  velocity structure seen across this large area. The two solid lines
  enclose a region where we calculated the velocity gradient along the
  direction indicated by the arrow; we found a peak velocity gradient
  $\sim$10~km~s$^{-1}$~pc$^{-1}$ along this direction. \textit{Right:}
  Zoom-in of the ridge zone centroid velocities. Between the solid
  white bars, we measured an average velocity difference
  $\sim$0.3~\kms{} perpendicular to the major axis of the
  filament. A 0.1~pc scale bar and
  synthesized beam is shown in the bottom right corner of each image.}
\label{fig:gradients}
\end{figure}

\begin{figure}[H]
\centering 
\includegraphics[scale=0.9]{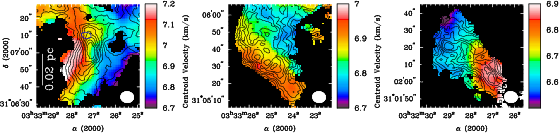}
\caption{\small Zoom-in on three small-scale regions with orderly
  centroid velocity fields, highlighting CLASSy's ability to resolve
  small-scale kinematic features, in addition to the large-scale
  structures seen in Figures~\ref{fig:vlsr} and
  \ref{fig:gradients}. The greyscale intensity is centroid velocity,
  and the contours are integrated intensity (contour levels in
  intervals of 10\% of the image peak intensity). The left image shows
  the dense gas around the compact continuum core toward Per-emb~30
  (the continuum core is identified with a single contour; colored
  blue in the online version). The center image is of gas south of
  Per-emb~30 in the main core zone, and the right image is of gas in
  the SW clumps zone. The synthesized beam is shown in the lower-right
  corner of each image, and a 0.02~pc scale bar is shown in the left
  image.}
\label{fig:order}
\end{figure}

Analyzing all the details of these kinematic maps is beyond the scope
of this paper. However, some features to point out are as
follows:

\begin{enumerate}

\item The most kinematic complexity exists in the main core zone---it
  has velocity gradients up to $\sim$10~\kmsppc \\(see
  Figure~\ref{fig:gradients}, left), and velocity dispersions ranging
  from 0.05~\kms{} all the way up to 0.5~\kms. In comparison, the gas
  structures in the ridge and SW clump zones show smaller variations
  in centroid velocity and velocity dispersion, and the velocity
  dispersions are consistently narrower than those found in the main
  core zone.

\item The narrow filament in the ridge zone, which can be seen in the
  all three kinematic maps, is redshifted by 1.5~\kms{} relative to
  the rest of the Barnard~1 emission. The mean velocity dispersion
  along the filament is 0.12~\kms{}, 0.16~\kms, and 0.20~\kms, for
  \NtwoH{}, HCN, and \HCO, respectively. At an assumed kinetic
  temperature of 11~K based on GBT data \citep{2008ApJS..175..509R},
  \NtwoH{} and HCN are detecting subsonic gas motions and \HCO{} is
  approaching the sonic speed. The 1.5~\kms{} radial velocity of the
  filament gas relative to the bulk Barnard~1 gas suggests that it did
  not simply fragment from the main reservoir of gas in the region. It
  is possible that a nearby flow of material piled onto the larger
  filament and flowed around its western edge; this would cause a
  column density increase along that edge, thereby strengthening the
  molecular emission and causing redshift. One problem with this
  scenario is that we might expect more turbulent linewidths for gas
  involved in a colliding flow. However, this type of flow event could
  have happened long enough ago that gravity had time to collect the
  gas into the filament we now see. Another possibility is that we are
  viewing a sheet-like structure edge on.

\item The largest structure in the ridge zone has a velocity gradient
  perpendicular to its major axis (see Figure~\ref{fig:gradients},
  right), which is a feature common in numerical simulations of
  filament formation from planar converging, turbulent flows
  \citep{2014ApJ...785...69C}.  This type of velocity gradient is seen
  in several CLASSy filaments across our five regions; Serpens
    South examples are highlighted in \citet{2014ApJ...790L..19F},
  and we are preparing a paper linking these observations with the
  numerical simulations (Mundy et al., in preparation).

\item Upon close inspection, several regions of the cloud have
  well-organized \NtwoH{} centroid velocity fields---for example, the
  dense gas around B1-c shows clear signatures of envelope rotation
  that was previously reported in \citet{2006ApJ...652.1374M}. We show
  other examples of orderly velocity fields in Figure~\ref{fig:order},
  coming from dense gas peaks that are spatially separated from other
  structures at low emission levels.  One of the examples is the dense
  gas surrounding the newly detected compact continuum core within
  Per-emb~30, while the other two have no detected protostars at their
  center. This highlights the ability of CLASSy data to probe the
  small-scale velocity structure around cores in addition to the
  large-scale velocity structure across the entire region.

\item The best-fit values of the centroid velocity and velocity
  dispersion toward the B1-c core are consistent with the results
  published by \citet{2006ApJ...652.1374M}. We detect the same
  evidence of a rotating \NtwoH{} envelope in the centroid velocity
  map, and detect the same increased velocity dispersion along the
  \HCO{} and HCN outflow axis. Also, the \NtwoH{} centroid velocities
  and velocity dispersions across the field match well with GBT
  NH$_{3}$ observations toward select dust cores
  \citep{2008ApJS..175..509R}.

\end{enumerate}

\section{DENDROGRAM ANALYSIS OF \NtwoH}
   
The previous two sections highlighted the dense gas morphology and
kinematics across Barnard~1. Here we quantitatively analyze the
\NtwoH{} position-position-velocity cube to create a census of dense
gas structures. We chose \NtwoH{} over HCN and \HCO{} because it is
our best tracer of the dense gas that is actively participating in
star formation---it closely mimics the dust emission and is less
  affected by absorption from lower density foreground gas than HCN
  and \HCO.

Several methods exist for identifying structures in an image or cube,
and the most appropriate method depends on the data and science
goals. In Appendix~A, we compare a widely used clump-finding algorithm
to the dendrogram method, and conclude that a dendrogram analysis is
more suitable for identifying resolved, dense gas structures in nearby
molecular clouds. A dendrogram tracks emission structure as a function
of isocontour level intensity, and represents the structure as a tree
hierarchy made up of leaves and branches \citep{1992ApJ...393..172H,
  2008ApJ...679.1338R, 2009Natur.457...63G}. Leaves are smaller-scale,
brighter objects at the top of the emission hierarchy that do not
break-up into further substructure, while branches are the
larger-scale, fainter objects lower in the hierarchy that do break-up
into substructure. The major benefit of a dendrogram analysis over a
clump-finding analysis rests in this ability to represent all of the
spatial scales in a dataset, as opposed to forcing all emission into
distinct clumps associated with a local peak.  See Appendix~A and
\citet{2009Natur.457...63G} for more discussion.

In Appendix B, we describe our modifications to the standard
dendrogram algorithm that enable non-binary hierarchies, and we argue
that non-binary dendrograms provide a more statistically meaningful
way to represent hierarchical emission in the presence of noise.  The
important modifications include: 1) restricting branching to intensity
steps equal to integer values of the 1-$\sigma$ sensitivity of the
data instead of allowing branching at infinitely small intensity
steps, and 2) using an algorithm that can cluster more than two
objects into a single group instead of being restricted to clustering
two objects at a time. These changes create an observable emission
hierarchy within the noise limits of the data, and allow the
quantification of the hierarchical complexity of a dendrogram using
tree statistics. The rest of this section focuses on our non-binary
dendrogram analysis of the \NtwoH{} gas in Barnard~1.

\subsection{The Non-binary Dendrogram}
 
We used our new non-binary dendrogram code to identify gas structures
traced by the isolated hyperfine component of \NtwoH{}.  We chose to
use the isolated hyperfine component of \NtwoH{} since it is
sufficiently separated from other components in velocity space to
prevent contaminating our object identification along the velocity
axis. Before running the data through the dendrogram code, we binned
the cube by two velocity channels to improve the signal-to-noise. No
velocity information was lost since there are no lines of sight that
contain two or more independent structures within 1.0~\kms{} of each
other. We found that binning by two channels provides the most
improvement to signal-to-noise without biasing the maps towards
wide-line regions.  The 1-$\sigma$ sensitivity of the binned data cube
is 0.094~\Jybm.

Our non-binary dendrogram code takes the same input parameters as the
standard IDL implementation discussed in
\citet{2008ApJ...679.1338R}. We ran it with the following critical
inputs and parameters: (1) a masked cube containing all pixels greater
than or equal to 4-$\sigma$ intensity, along with adjacent pixels of
at least 2.5-$\sigma$ intensity, (2) a set of local maxima greater
than or equal to all their neighbors in 10\arcsec{} by 10\arcsec{} by
three channel (0.94~\kms) spatial-velocity pixels, (3) a requirement
that a local maximum must peak 2-$\sigma$ above the intensity where it
first merges with another local maximum for it to be considered a
leaf, (4) a requirement of at least three synthesized beams of
spatial-velocity pixels for a leaf to be considered real.

\begin{figure}[H]
\centering \includegraphics[scale=0.2]{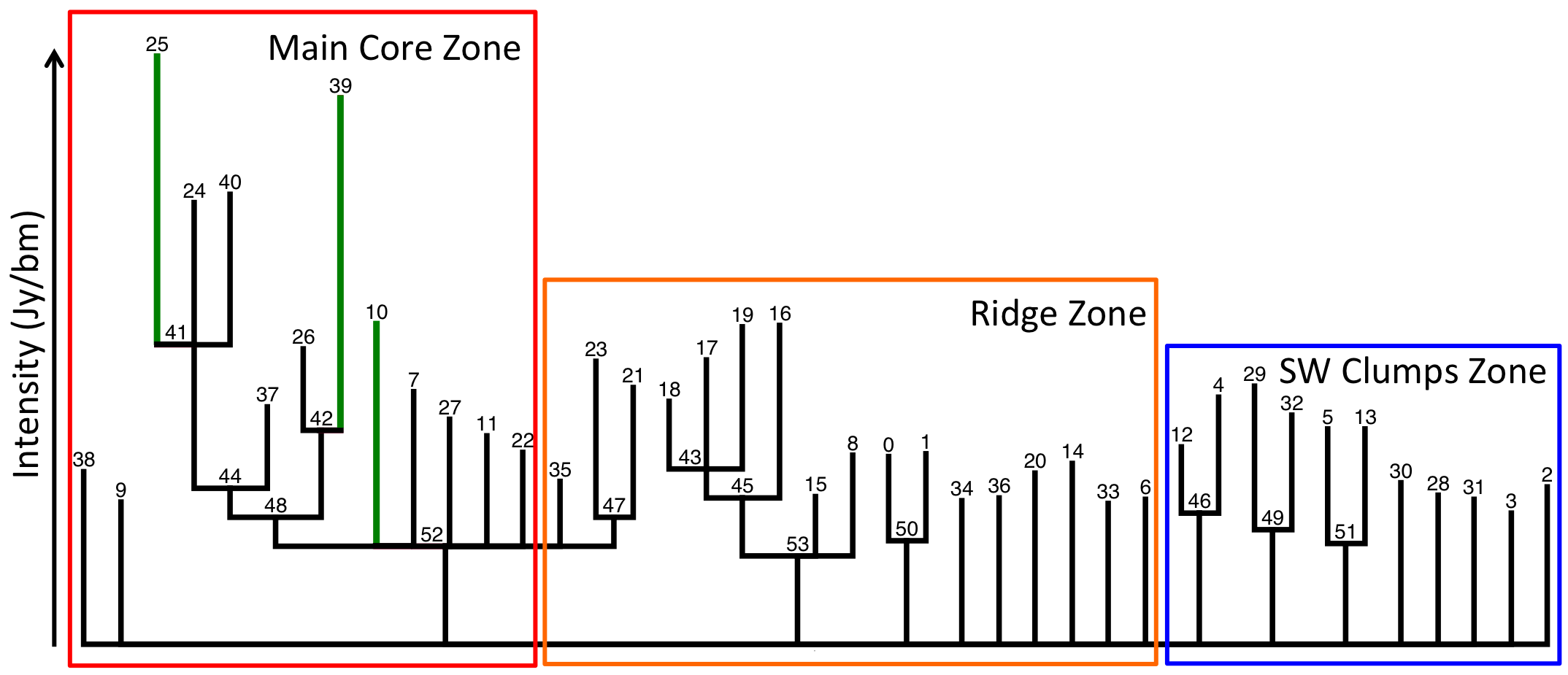}
\caption{\small The non-binary dendrogram for Barnard 1 is shown; it
  represents the hierarchical structure of the \NtwoH{} gas. There are
  41 leaves (numbered 0 through 40), and 13 branches (numbered 41
  through 53) on the dendrogram. The vertical axis represents the
  \Jybm{} intensity for a given location within the gas
  hierarchy. Branching occurs in integer multiples of the 1-$\sigma$
  sensitivity of the binned data cube used in this analysis. The
  horizontal axis is ordered according to zone, though the order
  within a given zone has no physical meaning. Leaves 10, 25, and 39
  (colored green in the online version) peak at least 6-$\sigma$ above
  their first merge level.}
\label{fig:b1tree}
\end{figure}

\begin{figure}[H]
\centering
\includegraphics[scale=0.55]{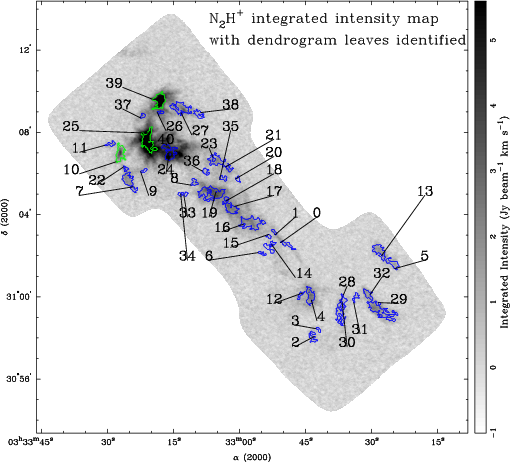}
\caption{ \small The \NtwoH{} integrated intensity map (\Jybmkms) is
  shown in greyscale. The overplotted contours (green and blue in the
  online version) are two-dimensional representations of the
  three-dimensional (PPV) leaf structures found in Barnard~1.  Green
  contours represent leaves with contrast greater than
  6-$\sigma$. Each leaf is labeled with its number that can be
  referenced to Figure~\ref{fig:b1tree} and Table~\ref{tbl:dendro}.}
\label{fig:leafmom0}
\end{figure}

\begin{figure}[H]
\centering
\includegraphics[scale=0.75]{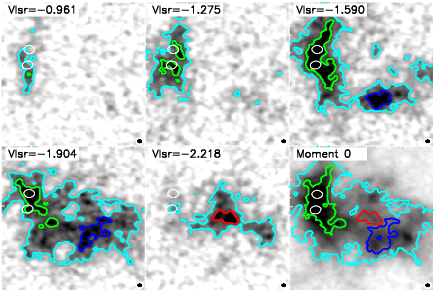}
\caption{ \small \NtwoH{} emission surrounding the B1-b double core,
  viewed across five binned velocity channels used in the dendrogram
  analysis, and across the moment zero map. For the channel maps, the
  greyscale represents the intensity of emission (\Jybm), and the
  colored contours (individually colored in the online version)
  represent the spatial extent of the dendrogram leaves and branches
  in that given channel. The single white contours in all maps
  represent the 4-$\sigma$ detection level of the B1-b continuum
  sources. In the online version, leaf 25 is green, 40 is red, and 24
  is blue, while branch 41 is cyan. For the moment zero map, the
  greyscale represents the integrated intensity of emission
  (\Jybmkms), and the colored contours represent two-dimensional
  representations of the three-dimensional dendrogram structures seen
  in the channel maps.}
\label{fig:dendrochannel}
\end{figure}

The non-binary dendrogram shown in Figure~\ref{fig:b1tree} contains 41
leaves and 13 branches. The vertical axis of the dendrogram
represents the intensity range of the pixels belonging to a leaf or
branch. The horizontal axis does not normally carry physical meaning,
but in this case we arranged the leaves and branches according to
zone. Figure~\ref{fig:leafmom0} shows two-dimensional representations
of all leaves overlaid on the \NtwoH{} integrated intensity map; these
representations were created by integrating over the velocity axis of
the leaves and contouring the maximal RA-DEC extent of the integrated
emission. The leaves that peak at least 6-$\sigma$ in intensity above
their first branch are colored green (leaves 10, 25, and 39).  Leaf 25
is the strongest, with a peak intensity of 1.91~\Jybm (in the binned
data cube), and represents the gas around B1-b. Leaf 39 is the next
strongest at 1.77~\Jybm, and represents gas around B1-c.  Leaf 10
peaks at 1.04~\Jybm, and represents gas around the newly detected
compact continuum source within Per-emb~30. As a reminder that the
identification of leaves and branches was done in three dimensions,
Figure~\ref{fig:dendrochannel} shows five \NtwoH{} velocity channels
near the B1-b cores with the isocontours of leaves 24, 25, and 40, and
branch 41, identified with distinct colors.

We are using the 6-$\sigma$ contrast criteria to highlight the
strongest leaves before we are able to confidently discuss the virial
boundedness of actual ``cores'' and ``clumps''---defining bound cores
requires accurate measurements of mass to go along with our
high-resolution kinematic information, which is beyond the scope of
this paper.  We chose a 6-$\sigma$ contrast to highlight the gas in
Barnard~1 that is located near existing compact continuum
objects---this gas is likely bound. We will apply the same contrast
cut to other CLASSy regions. While we cannot discuss boundedness of
individual gas cores at this stage, we can discuss the properties of
the strongest, highest-column density, dendrogram features and compare
them to weaker features in the field. Leaves with 6-$\sigma$ or
greater contrast will be referred to as high-contrast leaves later in
the paper, and the rest of the leaves will be called low-contrast
leaves.

The tree structure presented here captures the qualitative
hierarchical nature of the dense gas in Barnard 1, but it does not
give quantitative information about the hierarchy or physical
properties of the leaves and branches. Doing science with a dendrogram
requires extracting tree statistics and physical properties of the
leaves and branches, such as size, axis ratio, linewidth, mass,
luminosity, etc. We present three simple tree statistics and the
spatial and kinematic properties of leaves and branches in this paper.
In future papers, we will perform more detailed comparisons between
dendrogram properties across multiple CLASSy regions.

\subsection{Dendrogram Tree Statistics}

Computation of tree statistics from a non-binary dendrogram is one
method of quantifying a tree structure. \citet{1992ApJ...393..172H}
were the first to develop tree statistics for the astronomy community
in order to quantify the hierarchical nature of molecular cloud
structure (see their discussion of ``merged trees,'' which are
analogous to our non-binary trees). They describe several statistics
that can be computed from non-binary trees; we describe three of the
simplest statistics here.

The \textit{maximum branching level} is the integer number of
branching steps needed to reach the leaves at the top of the
hierarchy. A region that has undergone a lot of hierarchical
fragmentation will have more levels than a region that has not
fragmented, or that has fragmented in a single step into several
pieces with similar properties.  Branching levels are defined upward
from the base of the tree; the base (0~\Jybm{} intensity) is
considered level 0. An isolated leaf that sprouts directly from the
base is at level 0, while a leaf that grows from a branch one level
above the base is at level 1. The maximum branching level in our
Barnard~1 \NtwoH{} dendrogram is level 4, where leaves 24, 25, and 40
merge together into branch 41 (see Figure~\ref{fig:b1tree}). These
leaves are in the main core zone, which is the only zone with
detections of 3~mm compact continuum cores and Class 0/I YSOs. It is
reasonable that the largest amount of fragmentation has occurred in
the region with the most young star formation activity. That being
said, the ridge zone and SW clumps zone have maximum branching levels
of three and two, respectively, so we are dealing with small
variations of this statistic within Barnard~1.

A related tree statistic is the \textit{mean path length} of all
segments in a dendrogram. The path length of a segment is defined as
the number of branching steps required to go from a leaf to the bottom
of the tree (there are as many path lengths in a dendrogram as
  there are leaves). For example, the path length of leaf 25 is four,
while the path length of leaf 37 is three---the path length of leaf 30
is zero because it directly sprouts from the bottom of the tree. The
mean path length in the whole Barnard~1 \NtwoH{} dendrogram is 1.3
levels, while it is 2.4, 1.2, and 0.5 for the main core zone, ridge
zone, and SW clumps zone, respectively. Again, it is logical that the
main core zone, which is actively fragmenting into young stars, has
the largest mean path length.

It is important to point out that these statistics would be different
if we used the standard dendrogram algorithm instead of our non-binary
one. The standard algorithm forces binary branching by introducing
extra branching steps to ensure that only two objects merge at a
time. For example, leaves 24, 25, and 40 would not all be allowed to
merge into branch 41 even though that is what the data suggests is
happening---leaves 24 and 25 would be merged into one branch, which
would then be merged with leaf 40 into a second branch. This inflates
the path length of leaves 24 and 25, and increases the maximum
branching level of the tree. See Appendix B for more details about
differences between the standard dendrogram algorithm and the new
non-binary algorithm.

The third simple tree statistic is the \textit{mean branching ratio}
of all branches in a tree. The branching ratio for a single branch is
defined as the number of substructures into which it fragments
directly above it in the hierarchy. For example, the branching ratio
of branch 41 is three, because it has three leaves directly above
it. A very hierarchical region, where every object fragments into two
nested objects, will have a mean branching ratio of 2, while a region
that has fragmented in a single step into several pieces with similar
properties will have a much larger mean branching ratio.  The mean
branching ratio in our Barnard~1 \NtwoH{} dendrogram is 3.9.  We
include the branching ratio for the base of the tree in this
calculation, even though it is not explicitly defined with a branch
number -- this ensures that leaves that sprout directly from the base
are factored in. (Note that the branching ratio above the tree base is
always fixed at 2 for the standard dendrogram algorithm that forces
binary branching.)

These simple statistics provide measures of the amount and type of
fragmentation a given region has undergone; a region with a lot of
hierarchical fragmentation will have more levels, a larger mean path
length, and a smaller branching ratio, compared to a region that is
just beginning to form strong overdensities. A caveat is that the
absolute values of the branch statistics depend on algorithm
parameters and choice of allowed branching steps (reminder that we
restrict branching to intensity steps equal to integer values of the
1-$\sigma$ sensitivity of the data instead of allowing branching at
infinitely small intensity steps). However, the statistics can be
compared across trees as a differential measurement if the same
algorithm parameters and allowed branching steps are used; as pointed
out by \citet{1992ApJ...393..172H}, the power of tree statistics is
strongest when comparing regions with different star formation
properties. We will compare the Barnard~1 tree statistics with the
tree statistics of NGC~1333 and L1451 in an upcoming paper to explore
whether the hierarchical complexity of dense gas represented by tree
statistics correlates with the diversity of star formation activity in
the western half of Perseus.

Another caveat arises from linking dendrogram hierarchical complexity
to cloud fragmentation---tree statistics can be affected by projection
effects. All observations suffer from projection effects when
transforming the true position-position-position information of a
molecular cloud to observable position-position-velocity information
\citep[for an in-depth discussion of projection effects in spectral
  line data, see][]{2013ApJ...777..173B}. Projection effects distort
optically thick, space-filling emission more than optically thin
emission concentrated in one part of the cloud.  These effects are not
overly concerning in our data because the isolated hyperfine component
of \NtwoH{} is optically thin, and the \NtwoH{} emission is arising
only from the densest parts of the cloud.  If our data does suffer
from projection effects, we assume that our large mapping areas create
a big enough sample of dendrogram structures so that equal numbers of
projection effects occur in each region we will compare.

\subsection{Dendrogram Spatial Properties}

The size of an object is one of its most fundamental properties, but
defining the sizes of irregularly shaped objects in a uniform way is
not trivial. This is one challenge of high-angular-resolution surveys
of nearby molecular clouds---CLASSy resolves a rich morphology of
structures with a range of irregular shapes.

\begin{figure}[H]
\centering
\includegraphics[scale=0.75]{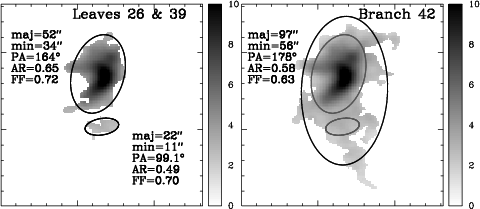}
\caption{ \small Example fits for the spatial properties of leaves and
  branches.  The left panel shows the \NtwoH{} integrated intensity
  (\Jybmkms) within the two-dimensional footprints of leaves 26 and
  39. We estimated the shape of these leaves with {\tt regionprops} in
  MATLAB; the fits are shown with the solid black ellipses, and the
  fit parameters are listed.  The axis ratio (AR) is the ratio of
  minor-to-major axis, and the filling factor (FF) is the number of
  greyscale pixels within the fitted ellipse divided by the total
  number of pixels within the ellipse. The right panel shows the
  \NtwoH{} integrated intensity within the footprint of branch 42,
  with its ellipse fit shown in black, and the ellipse fits of the
  leaves in grey. When fitting for the size of a branch, we include
  emission from the branch itself and all objects above it in the
  dendrogram---this results in a filled-in integrated intensity map
  for the fitting process, whereas using the emission from a single
  branch would give an ``onion-layer''-like structure.}
\label{fig:spaex}
\end{figure}

When fitting for the size of a leaf or branch, we consider all pixels
in its plane-of-sky footprint (as opposed to only considering the
``onion-layer'' emission of each individual branch, which is used to
derive kinematic properties in the next section---see
Figures~\ref{fig:spaex} and \ref{fig:kinex} for examples).  We use the
{\tt regionprops} program in MATLAB to fit an ellipse to the
integrated intensity footprint of each dendrogram object. All pixels
are given equal weight so that the ellipse is preferentially fit to
the largest scale of the object---this prevents strong emission at the
center of an object from driving the fit towards a smaller scale.  The
fit is defined with an RA centroid, DEC centroid, major axis, minor
axis, and position angle.  The location, major axis, minor axis, and
position angle are directly determined by the fit, and are listed in
columns 2-6 in Table~\ref{tbl:dendro}.  We do not report formal
uncertainties on these values since the spatial properties of
irregularly shaped objects are dependent on the chosen method, and we
do not want to mislead the reader as to the ``certainty'' of these
values.

To quantify the shape of an object, we use the axis ratio and filling
factor of the fit (columns 7 and 8 of Table~\ref{tbl:dendro}). The
axis ratio is defined as the ratio of the minor-to-major axis; a
circular fit will have an axis ratio of one, and a highly elongated
fit will have an axis ratio approaching zero. The filling factor is
defined as the area of emission within the fitted ellipse divided by
the area of the fitted ellipse; a circular object and an elliptical
object will both have a filling factor near one, but an irregular
object will likely have a filling factor much less than one,
regardless of the axis ratio of the fitted ellipse.  We lastly define
an object size (column 9 of Table~\ref{tbl:dendro}) as the geometric
mean of the major and minor axes.  The values reported in
Table~\ref{tbl:dendro} have been converted to parsecs, assuming a
distance of 235~pc.  Figure~\ref{fig:spaex} shows example spatial
property fitting results for leaves 26 and 39 and branch 42.

Histograms of the size, filling factor, and axis ratio for all leaves
and branches are plotted in the top row of
Figure~\ref{fig:histodendro}.  The trend in size is obvious---branches
represent larger structures than the leaves that peak within
them. Leaf sizes range from about 0.01~pc to 0.07~pc, while branch
sizes range from about 0.08~pc to 0.34~pc. The high-contrast leaves
are larger than the majority of low-contrast leaves. Nearly 25\% of
leaves have filling factors greater than 0.8, while all branches have
filling factors less than 0.8. This shows that leaves are better fit
as regular shapes compared to branches, but there is still a large
population of irregularly-shaped leaves that are far from round or
filamentary. The objects with the maximum axis ratios (leaf 14 and
branch 46) are paired with filling factors below 0.8, indicating that
they are not regularly shaped spheroids as the axis ratio alone might
be interpreted. There are no obvious distinctions between
high-contrast and low-contrast leaves in filling factor or axis ratio.

\subsection{Dendrogram Kinematic Properties}
One strength of CLASSy data is the kinematic information across a wide
range of spatial scales. We can use the integrated intensity
footprints of the dendrogram objects (leaf footprints shown in
Figure~\ref{fig:leafmom0}) in combination with the centroid velocity
and velocity dispersion maps in Figure~\ref{fig:vlsr} to determine the
kinematic properties of leaves and branches. The four properties in
Table~\ref{tbl:dendro} that we focus on are: the mean and rms centroid
velocity ($\langle \mathrm{V}_{\mathrm{lsr}} \rangle$ and
$\Delta$\vlsr, respectively), and the mean and rms velocity dispersion
($\langle \sigma \rangle$ and $\Delta \sigma$, respectively).

\begin{figure}[H]
\centering
\includegraphics[scale=0.75]{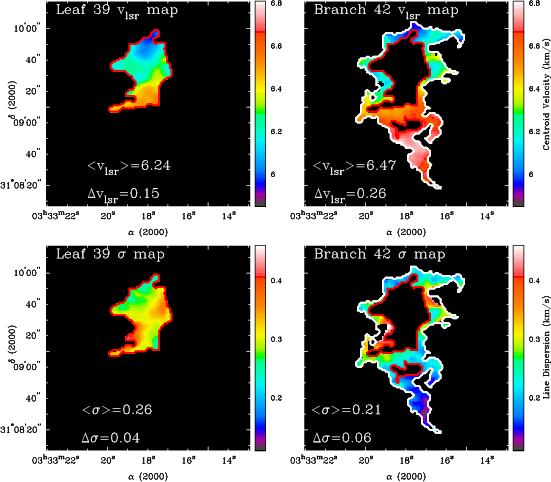}
\caption{ \small Examples of how we calculate the kinematic properties
  of leaves and branches, using the leaf emission towards B1-c, and
  the branch emission directly surrounding it. \textit{Upper left:}
  The integrated intensity footprint of leaf 39 is represented by the
  contour (colored red in the online version). The best fit centroid
  velocities (\vlsr) from Figure~\ref{fig:vlsr} are shown within that
  contour; to find the mean centroid velocity ($\langle
  \mathrm{V}_{\mathrm{lsr}} \rangle$) for leaf 39, we calculate the
  weighted mean of those pixels (weighing according to fit error at
  each pixel), and to find the rms centroid velocity ($\Delta$\vlsr)
  for leaf 39, we calculate the weighted standard deviation of those
  pixels. \textit{Lower left:} We show the best fit velocity
  dispersions ($\sigma$, not FWHM) from Figure~\ref{fig:vlsr} within
  the contour, and calculate the weighted mean velocity dispersion
  ($\langle \sigma \rangle$) and rms velocity dispersion ($\Delta
  \sigma$) as we did for the \vlsr{} values in the upper
  left. \textit{Upper right:} The integrated intensity footprint of
  branch 42 is represented by the outer contour (colored white in the
  online version), while those of leaf 39 and leaf 26 are represented
  by the inner contours (colored red in the online version). The best
  fit \vlsr{} from Figure~\ref{fig:vlsr} are shown only between the
  outer and inner contours. We are using information only at the
  branch spatial scales to calculate the kinematic properties of the
  branch---we exclude the kinematic properties of the leaves within
  the branch. As for the upper left figure, to find $\langle
  \mathrm{V}_{\mathrm{lsr}} \rangle$ for branch 42, we calculate the
  weighted mean of the shown pixels, and to find $\Delta$\vlsr{}, we
  calculate the weighted standard deviation of the shown
  pixels. \textit{Lower right:} The best fit $\sigma$ for branch 42
  are shown between the outer and inner contours.}
\label{fig:kinex}
\end{figure}

We illustrate how we derive these properties for leaves and branches
in Figure~\ref{fig:kinex}. For leaves, we mask the full \NtwoH{}
\vlsr{} and $\sigma$ maps with the integrated intensity footprint of
each leaf, and calculate the error weighted mean and standard
deviation of the masked pixels within the masked footprints. For
branches, we mask the full \NtwoH{} \vlsr{} and $\sigma$ maps with the
integrated intensity footprint of each branch, excluding pixels
associated with any leaves or branches inside the branch footprint. We
calculate the error weighted mean and standard deviation of the
remaining pixels belonging exclusively to the branch. By excluding
leaf pixels, we are calculating the kinematic properties of branches
only at the larger spatial scales that are not captured by the leaves
within them---this method allows us to parse kinematics across
different spatial scales, which enables a size-linewidth analysis in
the next section.  We only report kinematic properties of leaves and
branches in Table~\ref{tbl:dendro} that have three or more synthesized
beams worth of kinematic pixels; a low signal-to-noise feature may be
strong enough to be selected as a dendrogram object, but still too
weak to have enough kinematic data points based on our line fitting
criteria discussed in section~5.

\begin{figure}[H]
\centering
\includegraphics[scale=1]{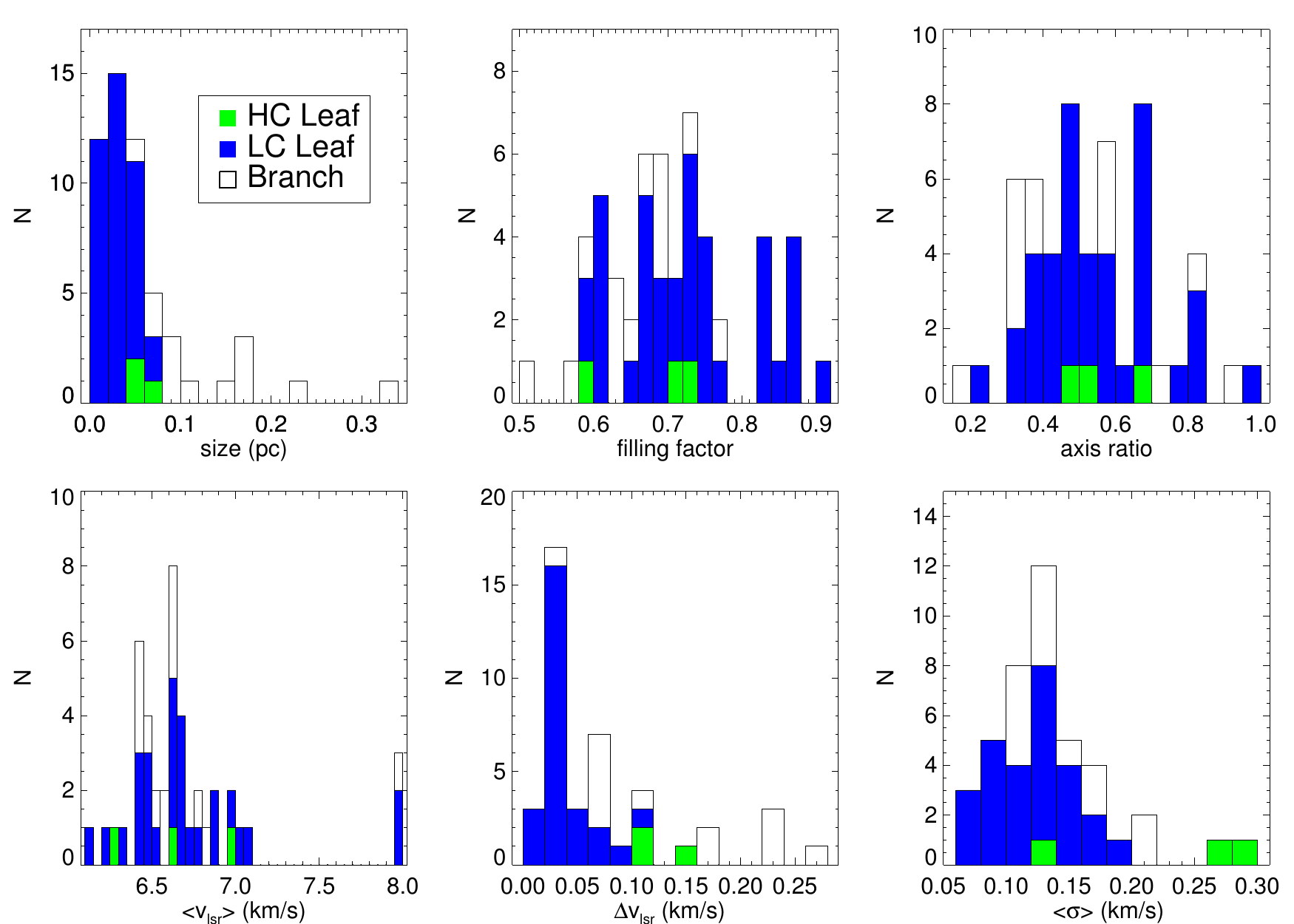}
\caption{ \small Histograms of basic dendrogram leaf and branch
  properties. High-contrast (HC) leaves, above 6-$\sigma$ contrast,
  are represented by green; low-contrast (LC) leaves, below 6-$\sigma$
  contrast, are represented by blue; branches are represented by
  white. See the text in sections~6.3 and 6.4 for a discussion of
  trends seen in these histograms.}
\label{fig:histodendro}
\end{figure}

Histograms of $\langle \mathrm{V}_{\mathrm{lsr}} \rangle$,
$\Delta$\vlsr, and $\langle \sigma \rangle$ are plotted in the bottom
row of Figure~\ref{fig:histodendro}. The distribution of $\langle
\mathrm{V}_{\mathrm{lsr}} \rangle$ shows that most of the gas in
Barnard 1 is near 6.5~\kms, while the gas from leaves 0 and 1 and
branch 50 (representing the narrow filament in the ridge zone) is
strongly redshifted.

There is a clear offset in the distribution of $\Delta$\vlsr{} between
leaves and branches: the majority of branches, which represent the
largest objects, have larger $\Delta$\vlsr{} than the
smaller leaves.  There are two effects that are likely causing this
trend.  The first effect is from organized velocity field motions
(e.g., indicative of rotation or feedback from outflows) that exist
towards a leaf and its surrounding branch.  For example, the \vlsr{}
field for the B1-c leaf and parent branch can be seen in
Figure~\ref{fig:kinex}. The branch that directly surrounds B1-c has a
larger $\Delta$\vlsr{} compared to the B1-c leaf due to
larger differences between the redshifted and blueshifted velocities
in the rotating branch.  However, the majority of leaf-branch pairs do
not exhibit matching, well-ordered velocity fields.  Therefore,
turbulence is the second likely explanation for this trend. Turbulence
has the property of causing scale-dependent spatial correlations
within a velocity field---points separated by larger distances will
have larger rms velocity differences between them than will points
that are closer together \citep{2007ARAA..45..565M}, meaning that
branches will have larger $\Delta$\vlsr{} values than
leaves.

The high-contrast leaves are preferentially shifted to larger
$\Delta$\vlsr{} relative to the low-contrast leaves.  The
high-contrast leaves all have known protostellar objects within them,
so their gas motions are most likely dominated by infall, rotation, or
feedback from outflows, all of which increase the variation in the
velocity field. The low-contrast leaves with the largest
$\Delta$\vlsr{} tend to have more organized centroid velocity patterns
(manifested as gradients) than the low-contrast leaves with the
smallest $\Delta$\vlsr{}, indicating that they may represent gas
around pre-stellar cores that are more likely to form stars in the
future.

The distribution of $\langle \sigma \rangle$ differs from the
distribution of $\Delta$\vlsr. The high-contrast leaves representing
the gas around B1-b and B1-c have the largest $\langle \sigma \rangle$
in the entire field, and the majority of branches have $\langle \sigma
\rangle$ that are similar to those of the majority of leaves. We would
naively expect that if the branches represent the largest objects
projected on the plane of the sky, then they should also represent the
most extended objects along our lines of sight, and should therefore
have the largest velocity dispersions \citep[assuming velocity
  dispersion scales proportionally with length-scale, which is the
  case for a turbulent cloud;][]{2007ARAA..45..565M}. We explore these
results in more detail in section~7, where we compare the sizes of
structures to their $\Delta$\vlsr{} and $\langle \sigma \rangle$ in
order to probe the physical and turbulent nature of Barnard~1.

\begin{deluxetable}{l l l c c c c c c c c c c c c c}
\tabletypesize{\tiny}
\tablecaption{\NtwoH{} Dendrogram Leaf and Branch Properties}
\tablewidth{0pt}
\setlength{\tabcolsep}{0.03in}
\tablehead
{
 \colhead{{No.}} & \colhead{{RA}} & \colhead{{DEC}} & \colhead{{Maj. Axis}} & \colhead{{Min. Axis}} & \colhead{{PA}} & \colhead{{Axis}} & \colhead{{Filling}} & \colhead{{Size}} & \colhead{{$\langle \mathrm{V}_{\mathrm{lsr}} \rangle$}} & \colhead{{$\Delta$\vlsr}} & \colhead{{$\langle \sigma \rangle$}} & \colhead{{$\Delta \sigma$}}  & \colhead{{Pk. Int.}}  & \colhead{{Contrast}}  & \colhead{{Level}}\\
 \colhead{{}} & \colhead{{(h:m:s)}} & \colhead{{(\deg:\arcmin:\arcsec)}} & \colhead{{(\arcsec)}} & \colhead{{(\arcsec)}} & \colhead{{(\deg)}} & \colhead{{Ratio}} & \colhead{{Factor}} & \colhead{{(pc)}} & \colhead{{(\kms)}} & \colhead{{(\kms)}} & \colhead{{(\kms)}} & \colhead{{(\kms)}} & \colhead{{(\Jybm)}} & \colhead{{}} & \colhead{{}}\\
 \colhead{{(1)}} & \colhead{{(2)}} & \colhead{{(3)}} & \colhead{{(4)}} & \colhead{{(5)}} & \colhead{{(6)}} & \colhead{{(7)}} & \colhead{{(8)}} & \colhead{{(9)}} & \colhead{{(10)}} & \colhead{{(11)}} & \colhead{{(12)}} & \colhead{{(13)}} & \colhead{{(14)}} & \colhead{{(15)}} & \colhead{{(16)}}

}
\startdata
\multicolumn{16}{c}{\bf{Leaves}}\\
\tableline
\noalign{\smallskip}
             0   &   03:32:49.1   &   +31:02:31.9   &          49.6   &          12.4   &          61.8   &          0.25   &          0.60   &         0.028   &           7.99(1)   &          0.02(0)   &          0.08(2)   &          0.04(1)   &          0.61   &           2.9    &    1   \\
             1   &   03:32:52.1   &   +31:03:09.3   &          22.9   &           8.3   &          41.1   &          0.36   &          0.84   &         0.016   &           7.98(1)   &          0.03(1)   &          0.08(1)   &          0.01(1)   &          0.62   &           3.0    &    1   \\
             2   &   03:32:43.4   &   +30:58:02.4   &          37.6   &          22.1   &          18.1   &          0.59   &          0.61   &         0.033   &           7.06(1)   &          0.02(1)   &          0.08(1)   &          0.01(0)   &          0.51   &           2.9    &    0   \\
             3   &   03:32:42.1   &   +30.58:24.3   &          19.9   &           9.1   &          36.7   &          0.46   &          0.82   &         0.015   &           ...   &          ...   &          ...   &          ...   &          0.42   &           2.0    &    0   \\
             4   &   03:32:43.9   &   +31:00:02.4   &          54.7   &          23.9   &           4.8   &          0.44   &          0.76   &         0.041   &           6.86(1)   &          0.03(0)   &          0.14(0)   &          0.02(0)   &          0.80   &           4.0    &    1   \\
             5   &   03:32:25.1   &   +31:01:36.9   &          48.5   &          20.0   &          55.1   &          0.41   &          0.68   &         0.035   &           6.99(1)   &          0.03(1)   &          0.12(1)   &          0.02(1)   &          0.70   &           4.0    &    1   \\
             6   &   03:32:54.9   &   +31:02:06.7   &          27.9   &           9.1   &          68.5   &          0.33   &          0.72   &         0.018   &           ...   &          ...   &          ...   &          ...   &          0.47   &           2.5    &    0   \\
             7   &   03:33:25.1   &   +31:05:36.7   &          66.7   &          27.8   &          30.7   &          0.42   &          0.67   &         0.049   &           6.88(1)   &          0.05(1)   &          0.07(0)   &          0.01(0)   &          0.82   &           5.3    &    1   \\
             8   &   03:33:10.3   &   +31:05:34.2   &          23.6   &          19.6   &          17.4   &          0.83   &          0.75   &         0.025   &           ...   &          ...  &          ...   &          ...   &          0.62   &           3.5    &    1   \\
             9   &   03:33:21.7   &   +31:06:07.8   &          22.9   &           8.9   &         118.2   &          0.39   &          0.82   &         0.016   &           ...   &          ...   &          ...   &          ...   &          0.46   &           2.4    &    0   \\
            10   &   03:33:27.1   &   +31:06:58.5   &          55.3   &          27.1   &         174.6   &          0.49   &          0.71   &         0.044   &           6.97(2)   &          0.10(1)   &          0.12(1)   &          0.03(0)   &          1.04   &           7.7    &    1   \\
            11   &   03:33:29.3   &   +31:07:26.1   &          32.5   &          13.4   &         101.7   &          0.41   &          0.68   &         0.024   &           7.02(1)   &          0.03(1)   &          0.10(1)   &          0.01(0)   &          0.68   &           3.8    &    1   \\
            12   &   03:32:46.2   &   +31:00:04.5   &          31.3   &          11.5   &         144.7   &          0.37   &          0.65   &         0.022   &           6.71(1)   &          0.03(1)   &          0.09(0)   &          0.01(0)   &          0.64   &           2.3    &    1   \\
            13   &   03:32:28.2   &   +31:02:13.1   &          53.1   &          27.7   &          37.2   &          0.52   &          0.74   &         0.044   &           6.66(2)   &          0.10(1)   &          0.10(1)   &          0.02(0)   &          0.70   &           4.0    &    1   \\
            14   &   03:32:53.1   &   +31:02:26.4   &          36.5   &          35.5   &          48.3   &          0.97   &          0.60   &         0.041   &           ...   &          ...   &          ...   &          ...   &          0.59   &           3.8    &    0   \\
            15   &   03:32:53.3   &   +31:02:55.7   &          16.6   &          11.4   &          55.9   &          0.69   &          0.83   &         0.016   &           ...   &          ...   &          ...   &          ...   &          0.48   &           2.1    &    1   \\
            16   &   03:32:57.3   &   +31:03:35.4   &          72.4   &          36.0   &          89.5   &          0.50   &          0.67   &         0.058   &           6.67(1)   &          0.04(0)   &          0.09(1)   &          0.03(0)   &          1.03   &           5.9    &    2   \\
            17   &   03:33:02.0   &   +31:04:18.4   &          49.0   &          26.2   &          55.7   &          0.53   &          0.72   &         0.041   &           6.61(1)   &          0.03(0)   &          0.15(0)   &          0.02(0)   &          0.92   &           3.8    &    3   \\
            18   &   03:33:03.0   &   +31:04:45.0   &          17.1   &          13.9   &          13.9   &          0.81   &          0.86   &         0.018   &           6.62(1)   &          0.02(1)   &          0.14(1)   &          0.01(0)   &          0.79   &           2.3    &    3   \\
            19   &   03:33:06.4   &   +31:05:02.2   &          80.3   &          40.1   &          88.8   &          0.50   &          0.67   &         0.065   &           6.64(1)   &          0.04(1)   &          0.13(1)   &          0.04(0)   &          1.03   &           4.9    &    3   \\
            20   &   03:33:00.3   &   +31:05:43.4   &          19.3   &          13.2   &          23.7   &          0.68   &          0.82   &         0.018   &           ...   &          ...   &          ...   &          ...   &          0.56   &           3.4    &    0   \\
            21   &   03:33:02.4   &   +31:06:15.5   &          25.8   &          17.0   &          38.6   &          0.66   &          0.68   &         0.024   &           6.67(1)   &          0.03(1)   &          0.13(1)   &          0.03(1)   &          0.83   &           4.5    &    2   \\
            22   &   03:33:25.9   &   +31:06:13.5   &          17.7   &          14.5   &          33.4   &          0.82   &          0.73   &         0.018   &           6.78(1)   &          0.02(1)   &          0.08(1)   &          0.01(1)   &          0.62   &           3.2    &    1   \\
            23   &   03:33:05.3   &   +31:06:38.3   &          51.4   &          33.5   &          69.5   &          0.65   &          0.60   &         0.047   &           6.60(1)   &          0.03(0)   &          0.11(0)   &          0.02(0)   &          0.92   &           5.4    &    2   \\
            24   &   03:33:15.7   &   +31:06:56.4   &          45.2   &          26.9   &         142.3   &          0.60   &          0.71   &         0.040   &           6.47(2)   &          0.10(2)   &          0.19(1)   &          0.03(0)   &          1.44   &           4.9    &    4   \\
            25   &   03:33:20.8   &   +31:07:33.0   &          72.7   &          38.7   &          11.4   &          0.53   &          0.58   &         0.060   &           6.62(2)   &          0.11(1)   &          0.29(1)   &          0.05(1)   &          1.91   &           9.9    &    4   \\
            26   &   03:33:18.0   &   +31:08:57.9   &          22.0   &          10.7   &          99.1   &          0.49   &          0.70   &         0.017   &           6.66(2)   &          0.03(1)   &          0.18(1)   &          0.02(1)   &          0.96   &           2.8    &    3   \\
            27   &   03:33:13.4   &   +31:09:10.2   &          61.2   &          32.8   &          73.1   &          0.54   &          0.66   &         0.051   &           6.51(1)   &          0.07(1)   &          0.17(1)   &          0.03(0)   &          0.73   &           4.4    &    1   \\
            28   &   03:32:37.1   &   +30.59:10.1   &          69.4   &          27.3   &           3.8   &          0.39   &          0.61   &         0.050   &           6.47(1)   &          0.04(1)   &          0.11(0)   &          0.02(0)   &          0.48   &           2.6    &    0   \\
            29   &   03:32:27.7   &   +30:59:18.2   &         101.8   &          34.5   &          59.8   &          0.34   &          0.59   &         0.068   &           6.44(1)   &          0.08(1)   &          0.16(0)   &          0.03(0)   &          0.84   &           5.0    &    1   \\
            30   &   03:32:36.4   &   +30:59:51.1   &          29.1   &          18.9   &          13.2   &          0.65   &          0.67   &         0.027   &           ...  &          ...   &          ...   &          ...   &          0.52   &           3.0   &    0   \\
            31   &   03:32:33.6   &   +30:59:59.1   &          32.0   &          15.4   &         151.7   &          0.48   &          0.58   &         0.025   &           ...   &         ...   &          ...   &          ...   &          0.47   &           2.5    &    0   \\
            32   &   03:32:30.8   &   +31:00:07.1   &          39.6   &          18.7   &          38.9   &          0.47   &          0.73   &         0.031   &           6.41(1)   &          0.04(1)   &          0.11(1)   &          0.03(0)   &          0.75   &           4.0    &    1   \\
            33   &   03:33:12.2   &   +31:04:59.9   &          15.4   &          10.3   &         144.5   &          0.67   &          0.87   &         0.014   &           ...  &          ...  &           ...   &       ...   &      0.46   &           2.3    &    0   \\
            34   &   03:33:13.4   &   +31:04:59.9   &          17.2   &          11.7   &         102.1   &          0.68   &          0.90   &         0.016   &           ...   &         ...   &          ...   &          ...   &          0.47   &           2.5    &    0   \\
            35   &   03:33:03.6   &   +31:05:46.4   &          25.6   &          14.7   &          75.3   &          0.57   &          0.72   &         0.022   &           6.42(2)   &          0.03(1)   &          0.13(1)   &          0.03(1)   &          0.53   &           2.2    &    1   \\
            36   &   03:33:07.7   &   +31:06:05.4   &          24.4   &          16.5   &          39.8   &          0.68   &          0.75   &         0.023   &           ...   &         ...   &          ...   &          ...   &          0.48   &           2.6    &    0   \\
            37   &   03:33:22.0   &   +31:08:53.6   &          16.5   &          13.2   &         139.6   &          0.80   &          0.87   &         0.017   &           6.32(2)   &          0.03(1)   &          0.14(1)   &          0.01(0)   &          0.77   &           2.8    &    3   \\
            38   &   03:33:09.2   &   +31:08:53.3   &          34.7   &          19.7   &          52.5   &          0.57   &          0.75   &         0.030   &           6.46(1)   &          0.03(1)   &          0.13(2)   &          0.04(1)   &          0.56   &           3.5    &    0   \\
            39   &   03:33:18.2   &   +31:09:32.2   &          52.1   &          34.0   &         164.4   &          0.65   &          0.72   &         0.048   &           6.24(3)   &          0.15(2)   &          0.26(1)   &          0.04(1)   &          1.77   &          11.5    &    3   \\
            40   &   03:33:16.7   &   +31:07:17.3   &          29.1   &          13.7   &          79.2   &          0.47   &          0.87   &         0.023   &           6.11(1)   &          0.04(1)   &          0.15(1)   &          0.02(0)   &          1.46   &           5.2    &    4   \\
\noalign{\smallskip}
\tableline
\noalign{\smallskip}
\multicolumn{16}{c}{\bf{Branches}}\\
\tableline
\noalign{\smallskip} 
            41   &   03:33:18.2   &   +31:07:20.6   &         178.0   &         103.1   &          70.6   &          0.58   &          0.66   &         0.154   &           6.45(2)   &          0.24(1)   &          0.21(0)   &          0.06(0)   &          0.97   &           ...    &    3   \\
            42   &   03:33:17.9   &   +31:09:21.3   &          96.6   &          56.0   &         178.0   &          0.58   &          0.63   &         0.084   &           6.47(4)   &          0.26(3)   &          0.21(1)   &          0.06(1)   &          0.69   &           ...    &    2   \\
            43   &   03:33:04.6   &   +31:04:45.5   &         153.5   &          55.3   &          60.5   &          0.36   &          0.59   &         0.105   &           6.64(1)   &          0.04(0)   &          0.14(1)   &          0.04(0)   &          0.57   &           ...    &    2   \\
            44   &   03:33:18.1   &   +31:07:23.5   &         197.6   &         116.9   &          67.0   &          0.59   &          0.68   &         0.173   &           6.51(2)   &          0.17(1)   &          0.17(1)   &          0.05(0)   &          0.51   &           ...    &    2   \\
            45   &   03:33:01.9   &   +31:04:23.6   &         249.6   &          84.1   &          55.6   &          0.34   &          0.69   &         0.165   &           6.63(1)   &          0.06(0)   &          0.12(0)   &          0.04(0)   &          0.47   &           ...    &    1   \\
            46   &   03:32:44.5   &   +31:00:00.4   &          79.2   &          75.1   &          81.4   &          0.95   &          0.76   &         0.088   &           6.80(1)   &          0.07(1)   &          0.12(0)   &          0.03(0)   &          0.43   &           ...    &    0   \\
            47   &   03:33:04.9   &   +31:06:38.4   &         107.1   &          38.1   &          50.3   &          0.36   &          0.63   &         0.073   &           6.59(1)   &          0.06(1)   &          0.12(1)   &          0.03(1)   &          0.41   &           ...    &    1   \\
            48   &   03:33:17.9   &   +31:07:47.8   &         237.1   &         177.5   &          17.5   &          0.75   &          0.64   &         0.234   &           6.54(2)   &          0.24(2)   &          0.16(1)   &          0.06(0)   &          0.41   &           ...    &    1   \\
            49   &   03:32:28.3   &   +30:59:29.1   &         148.3   &          47.8   &          45.1   &          0.32   &          0.69   &         0.096   &           6.44(1)   &          0.07(1)   &          0.14(1)   &          0.04(0)   &          0.37   &           ...    &    0   \\
            50   &   03:32:46.4   &   +31:02:45.5   &         103.4   &          18.9   &          47.8   &          0.18   &          0.56   &         0.050   &           7.99(2)   &          0.07(1)   &          0.12(4)   &          0.13(3)   &          0.34   &           ...    &    0   \\
            51   &   03:32:27.0   &   +31:01:58.4   &         115.7   &          38.2   &          47.6   &          0.33   &          0.63   &         0.076   &           6.70(5)   &          0.17(4)   &          0.11(1)   &          0.03(1)   &          0.33   &           ...    &    0   \\
            52   &   03:33:17.1   &   +31:07:34.2   &         324.8   &         267.3   &          73.1   &          0.82   &          0.50   &         0.336   &           6.67(2)   &          0.23(1)   &          0.12(0)   &          0.05(0)   &          0.32   &           ...    &    0   \\
            53   &   03:33:01.8   &   +31:04:21.9   &         274.0   &          90.2   &          54.4   &          0.33   &          0.72   &         0.179   &           6.63(2)   &          0.10(1)   &          0.12(1)   &          0.04(0)   &          0.29   &           ...    &    0   \\
\enddata
\vspace{-0.5cm} \tiny \tablecomments{(2)--(6) The position, major
  axis, minor axis, and position angle were determined from {\tt
    regionprops} in MATLAB. We do not report formal uncertainties of
  these values since the spatial properties of irregularly shaped
  objects is dependent on the chosen method.  \\ (7) Axis ratio,
  defined as the ratio of the minor axis to the major axis.\\ (8)
  Filling factor, defined as the area of the leaf or branch inscribed
  within the fitted ellipse, divided by the area of the fitted
  ellipse. \\ (9) Size, defined as the geometric mean of the major and
  minor axes, for an assumed distance of 235~pc. \\ (10) The weighted
  mean V$_{\mathrm{lsr}}$ of all fitted values within a leaf or
  branch. Weights are determined from the statistical uncertainties
  reported by the IDL MPFIT program. The error in the mean is reported
  in parentheses as the uncertainty in the last digit. It was computed
  as the standard error of the mean, $\Delta
  \mathrm{V}_{\mathrm{lsr}}/\sqrt N$, where $\Delta$\vlsr{} is the
  value in column 11 and $N$ is the number of beams' worth of pixels
  within a given object.  We report kinematic properties only for
  objects that have at least three synthesized beams' worth of
  kinematic pixels.  \\ (11) The weighted standard deviation of all
  fitted V$_{\mathrm{lsr}}$ values within a leaf or branch. The error
  was computed as the standard error of the standard deviation,
  $\Delta \mathrm{V}_{\mathrm{lsr}}/\sqrt{2(N-1)}$, assuming the
  sample of beams was drawn from a larger sample with a Gaussian
  velocity distribution. \\ (12) The weighted mean velocity dispersion
  of all fitted values within a leaf or branch. The error was computed
  as the standard error of the mean, $\Delta\sigma/\sqrt N$.  \\ (13)
  The weighted standard deviation of all fitted velocity dispersion
  values within a leaf or branch. The error was computed as the
  standard error of the standard deviation,
  $\Delta\sigma/\sqrt{2(N-1)}$.  \\ (14) For a leaf, this is the peak
  intensity measured in a single channel of our 2-channel binned
  dataset used in the dendrogram analysis. For a branch, this is the
  intensity level where the structures above it merge together.
  \\ (15) ``Contrast'' is defined as the difference between the peak
  intensity of a leaf and the height of its closest branch in the
  dendrogram, divided by the 1-$\sigma$ sensitivity of the
  data. \\ (16) The branching level in the dendrogram. For example,
  the base of the tree is level 0, so an isolated leaf that grows
  directly from the base is considered to be at level 0. A leaf that
  grows from a branch one level above the base will be at level 1,
  etc. \\}
\label{tbl:dendro}
\end{deluxetable}

\section{A CONNECTION BETWEEN SIZE-LINEWIDTH RELATIONS AND CLOUD STRUCTURE}

This paper presented the morphology and kinematics of the dense gas in
Barnard~1, along with a dendrogram analysis of the \NtwoH{}
emission. We conclude with an analysis of the spatial and kinematic
properties of dendrogram structures, and what they can tell us about
the turbulent and physical nature of the Barnard~1 cloud.

A turbulent cloud will have scale-dependent spatial correlations of
its velocity field that take on power law forms over a wide enough
range of spatial scales \citep{2007ARAA..45..565M}.  
Size-linewidth relations are commonly used to probe turbulence; there
have been numerous studies of observed and simulated molecular clouds
using a wide array of observational and statistical techniques, such
as correlation plots \citep{1981MNRAS.194..809L, 1987ApJ...319..730S},
structure function and autocorrelation
analysis \citep{1994ApJ...429..645M, 2002AA...390..307O}, principal
component analysis
\citep{2004ApJ...604..196B,2004ApJ...615L..45H}, and dendrograms
\citep{2008ApJ...679.1338R}, to name a few. The overall result is that most galactic
molecular clouds exhibit power-law scaling relations consistent with
turbulence in a compressible medium, where supersonic motions and
overlapping shocks are important (so-called ``Burgers
turbulence''). It is well accepted from these analyses that molecular
clouds are turbulent, but there are new insights that can be gained by
using size-linewidth relations derived from high angular resolution
observations.

\subsection{Total Linewidth of Dendrogram Structures}

Our dendrogram analysis identified \NtwoH{} structures in the
Barnard~1 hierarchy as either leaves or branches, and we evaluated the
size and kinematic properties of those objects in section~6.  Since
the spatial scales of dendrogram structures in Barnard~1 range from
$\sim$0.01-0.3~pc, we can utilize these structures to investigate the
turbulent properties of the cloud. Before discussing size-linewidth
relations that use the full angular resolution of our observations, it
is useful to consider a relation that uses the ``total'' linewidth of
each dendrogram structure -- this relation can be compared with
classical size-linewidth relations from \citet{1981MNRAS.194..809L}
and \citet{1987ApJ...319..730S}, in which the detailed kinematic
variation within individual clouds was not resolved.  We calculated a
total linewidth for each structure by summing all the spectra assigned
to it and fitting the velocity dispersion of the resulting spectrum
(the $\sigma$, not FWHM, of the line); for a visualization, see
Figure~\ref{fig:kinex}, which shows the kinematic maps of an example
leaf and branch. In that figure, all the spectra within the red
contour of leaf 39, and all the spectra within the white contours of
branch 42 (including the leaf emission), are summed to calculate the
total linewidths of those respective structures.

\begin{figure}[H]
\centering
\includegraphics[scale=0.62]{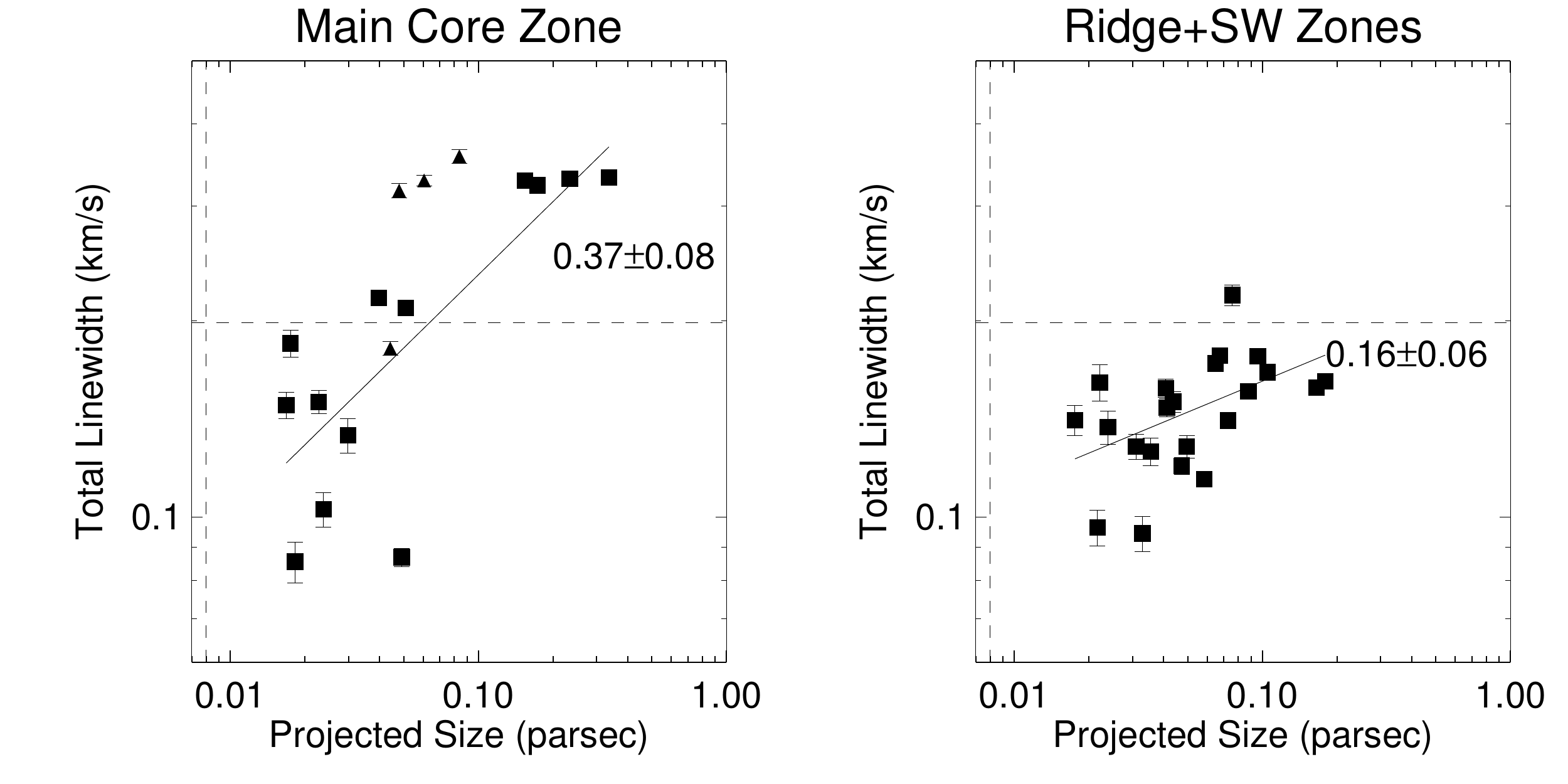}
\caption{ \small 
Scaling relations between projected structure size and total linewidth
for the main core zone and the combined ridge and SW clumps zones.
The projected size for each dendrogram structure is defined as the
geometric mean of its best-fit major and minor axes. The majority of
structures are labeled with squares, while the three leaves toward the
three strongest compact continuum detections in main core zone are
marked with triangles, along with the branch directly surrounding the
B1-c leaf. The solid line represents a single power-law fit to the
square points. The horizontal line represents the typical thermal
speed for H$_{2}$ at gas kinetic temperatures near 11~K. The vertical
line represents our spatial resolution of $\sim$0.008~pc.}
\label{fig:global}
\end{figure}

Figure~\ref{fig:global} shows plots of projected size versus total
linewidth for all dendrogram structures that have kinematic values
listed in Table~\ref{tbl:dendro}, excluding the objects associated
with the peculiar narrow filament in the ridge zone (leaves 0 and 1,
branch 50). We separate structures in the main core zone (MCZ) from
structures in the ridge and SW clumps zones (RSWZ) because the main
core zone has a cluster of protostars driving outflows. Keeping the
zones separate allows us to see whether or not they have different
turbulent or physical properties.  We also separately identify the few
structures surrounding compact continuum cores, since their
non-thermal gas motions are likely influenced by their central source.

The data in Figure~\ref{fig:global} are scattered and could be fit by
several functions. However, there does appear to be variation of total
linewidth with size in both zones, with the linewidth varying more
strongly with size in the MCZ.  If we use a single power-law fit, the
MCZ correlation is best fit with a slope of 0.37~$\pm$~0.08, and the
RSWZ correlation slope is 0.16~$\pm$~0.06.  The steeper slope in the
MCZ could be due to outflows adding energy to the turbulence at these
scales.  The best-fit slope from the MCZ is consistent with the result
\citep{1981MNRAS.194..809L} that clouds ranging in size from
0.1 to 100~pc have a power-law slope of 0.38.  However, we are probing
the scaling relation across much smaller spatial scales than has
typically been done, and unlike previous studies, we are using
high-density molecular tracers that are sensitive to non-thermal
motions near the sonic scale of the cloud. Therefore, directly
comparing our results to studies that used larger scales and
lower-density gas tracers, such as $^{12}$CO, should be considered
with caution.

\subsection{Resolved Linewidths of Dendrogram Structures}

We have two kinematic measurements of leaves and branches in
Table~\ref{tbl:dendro} that use the full angular resolution of our
dataset -- they complement the total linewidth and enable a different
type of size-linewidth analysis.  The first measurement is the \vlsr{}
variation, $\Delta$\vlsr. For a leaf or branch of size L,
$\Delta$\vlsr(L) is computed by measuring the rms variation of the
(beam-scale) centroid velocity over the whole structure.  The second
measurement is the mean non-thermal velocity dispersion, $\langle
\sigma \rangle_{\mathrm{nt}}$.  To calculate $\langle \sigma
\rangle_{\mathrm{nt}}$ from $\langle \sigma \rangle$
Table~\ref{tbl:dendro}, we remove the thermal velocity dispersion,
$\sigma_{\mathrm{th}}$, of 11~K \NtwoH{} gas, which is the typical gas
kinetic temperature across the entire Barnard~1 region
\citep{2008ApJS..175..509R}: $\langle \sigma \rangle_{\mathrm{nt}} =
\sqrt[]{\langle \sigma \rangle^{2} - \sigma_{\mathrm{th}}^{2}}$.
Recall that $\langle \sigma \rangle$ for a structure is defined by
averaging all the individual velocity dispersions over the structure;
each individual pointing captures the velocity dispersion along the
line-of-sight for a given beam-width.

\begin{figure}[]
\centering
\includegraphics[scale=0.8]{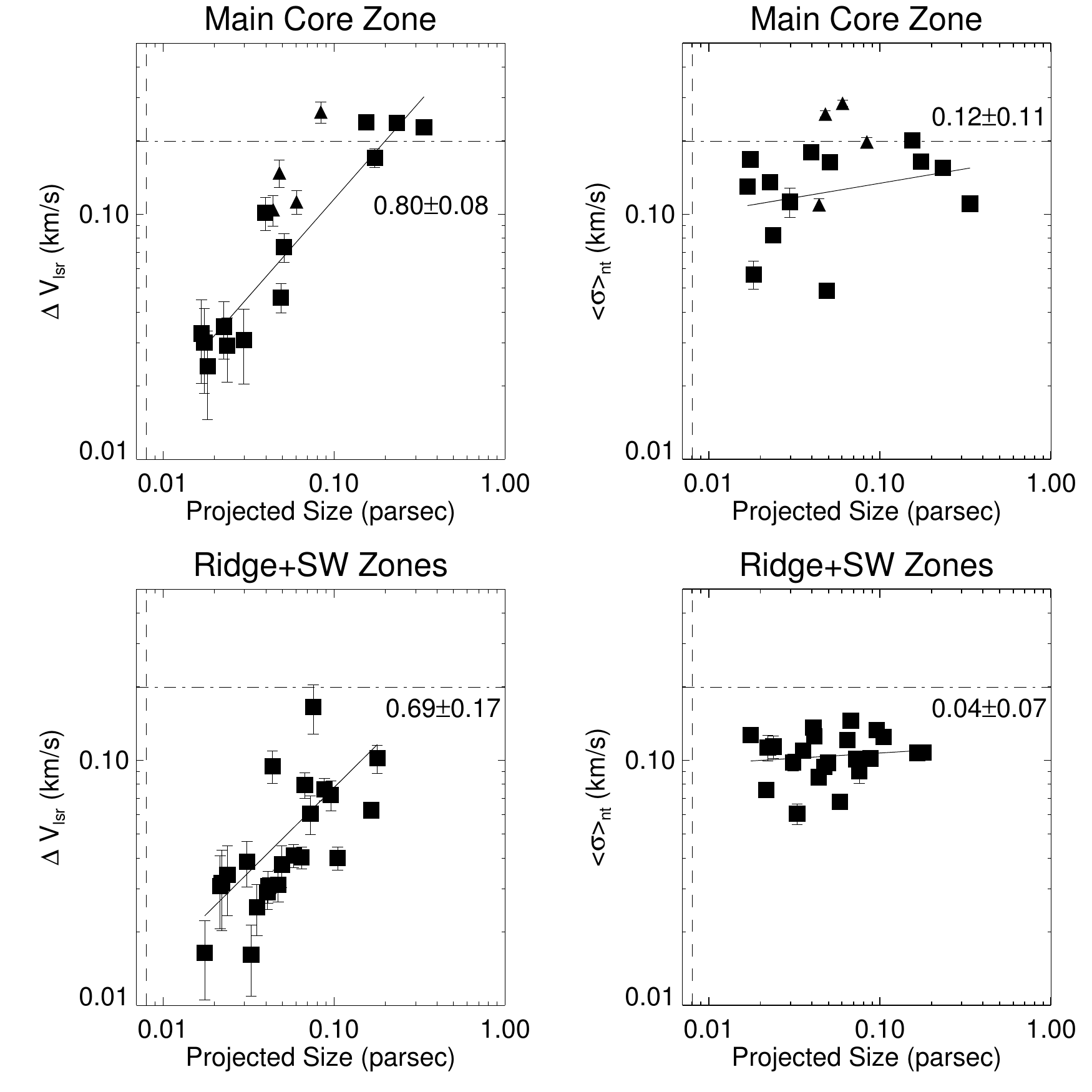}
\caption{ 
\small 
\textit{Left}: Scaling
  relations between projected structure size and \vlsr{} variation
  ($\Delta$\vlsr) for the main core zone and the combined
  ridge and SW clumps zones.
\textit{Right}: Scaling relations between projected structure size and
mean non-thermal velocity dispersion
($\langle \sigma \rangle_{\mathrm{nt}}$) for the main core zone and
the combined ridge and SW clumps zones. The solid line represents a
single power-law fit to the square points. The horizontal line
represents the typical thermal speed for H$_{2}$ at gas kinetic
temperatures near 11~K. The vertical line represents our spatial
resolution of $\sim$0.008~pc.}
\label{fig:decon}
\end{figure}

We compare the dependences of $\Delta$\vlsr{} and $\langle \sigma
\rangle_{\mathrm{nt}}$ on projected structure size in
Figure~\ref{fig:decon}. Size versus $\Delta$\vlsr{} for the dendrogram
structures in both zones is plotted on the left; the data appears to
follow a power-law relation. There is a steep correlation in both
regions, with power-law slopes of 0.80~$\pm$~0.08 in the MCZ and
0.69~$\pm$~0.17 in the RSWZ. The value of $\Delta$\vlsr{} is very low
for the smallest objects in all zones, and rises to the sound speed
for the largest objects in the MCZ.  Size versus $\langle \sigma
\rangle_{\mathrm{nt}}$ is plotted on the right side
Figure~\ref{fig:decon}.  There is a much flatter correlation between
size and $\langle \sigma \rangle_{\mathrm{nt}}$ in both zones, with a
power-law slope of 0.12~$\pm$~0.11 in the MCZ and 0.04~$\pm$~0.07 in
the RSWZ.  The mean non-thermal velocity dispersions are all sub-sonic
to trans-sonic in the MCZ (though still superthermal).  The two
structures with largest $\langle \sigma \rangle_{\mathrm{nt}}$ are
surrounding B1-b and B1-c; it is likely that the non-thermal motions
of the gas around them is boosted by rotation and/or outflows and is
not purely turbulent.  The $\langle \sigma \rangle_{\mathrm{nt}}$ are
all sub-sonic in the RSWZ.

Why are the dependences of $\langle \sigma \rangle_{\mathrm{nt}}$ and
$\Delta$\vlsr{} on projected structure size so different?
In the following section, we set up a theoretical framework that can
explain this disparity in terms of differences between a structure's
projected size and its depth into the plane of the sky.

\subsection{Inferring Cloud Depth from Resolved Size-Linewidth Relations}

In this section, we assume that the observed non-thermal gas motions
in our dendrogram structures are generated by isotropic,
three-dimensional turbulence. In the previous section, we noted a few
structures surrounding compact continuum cores that likely have
non-thermal gas motions due to the influence of their central source;
we exclude them from this analysis.

The total turbulent linewidth of a structure is most strongly
  influenced by the largest scale an observation encompasses, under
the assumptions that the gas we observe is isotropically turbulent,
and optically thin with uniform excitation conditions. If the
structure is wider across the plane of the sky than it is deep into
the sky, then its turbulent linewidth will be most strongly
  influenced by its projected size. But if the structure is deeper
into the sky than in either of the projected directions, then its
turbulent linewidth will be most strongly influenced by its
depth.

For a structure that is many resolution elements across (the case for
all of our dendrogram structures), $\Delta$\vlsr{} and
$\langle \sigma \rangle_{\mathrm{nt}}$ probe turbulence in different
ways from the total linewidth. The value of \vlsr{} on any given
line-of-sight is a measure of the mean motion of gas along that
line-of-sight, which varies from one line-of-sight to another
line-of-sight as a consequence of turbulence.  The \vlsr{} variation,
$\Delta$\vlsr, is expected to increase with the projected
scale of a structure if turbulence increases with scale, as is the
case for the observed and simulated turbulent power spectra in
molecular clouds.  For any given line-of-sight, the turbulent
contribution to $\langle \sigma \rangle$ will be set by the larger of
the beam size and the structure depth.  Since our beam size is very
small ($\sim$0.008 pc), we expect that the linewidth along individual
lines-of-sight will be dominated by structure depth, and thus
$\langle \sigma \rangle_{\mathrm{nt}}$ for a structure will depend
primarily on the mean (unseen) depth of that structure into the sky.
Statistically, along many lines-of-sight for a resolved structure, the
comparison of linewidths set by the projected size of the structure
($\Delta$\vlsr) to linewidths set by the depth of the
structure ($\langle \sigma \rangle_{\mathrm{nt}}$) should allow an
estimation of the structure depth.

We can use a mathematical framework to explain how we can compare
$\Delta$\vlsr{} to $\langle \sigma \rangle_{\mathrm{nt}}$ to learn
about the typical depth of gas structures. In the
full three dimensional turbulence, line-of-sight turbulent velocities
($v_{z,\mathrm{turb}}$) can be written as
\begin{equation}
v_{z,\mathrm{turb}}(x,y,z)=\sum_{k_x,k_y,k_z}
v_{p}(k_x,k_y,k_z) \exp[i (k_x x + k_y y +k_z z)],
\label{eq:turb}
\end{equation}
where $v_{p}(k_x,k_y,k_z)$ is the turbulent velocity power spectrum,
$k_{x}=(2 \pi/L_{x})n_{x}$, $k_{y}=(2 \pi/L_{y})n_{y}$, and $k_{z}=(2
\pi/L_{z})n_{z}$. $L_{x}$ is the size of an observed structure in the
$x$ direction, and $n_{x}$ is the number of full waves of a given
turbulent velocity component that fit in the $x$-direction (the same
is true for the $y$ and $z$ components in these equations).  If the
turbulent motions are Gaussian and isotropic, then
$v_{p}(k_{x},k_{y},k_{z})$ is drawn from a normal distribution whose
standard deviation depends only on the magnitude of the $k$-vector,
$\vert {\bf k} \vert =\sqrt{k_{x}^2+k_{y}^2+k_{z}^2}$.  We assign the
$z$ direction as the line-of-sight, so that in Equation~\ref{eq:turb},
$v_{z,\mathrm{turb}}(x,y,z)$ is the $z$-component of the turbulence,
and there are other independent components in the $x$ and $y$
directions.

\begin{figure}[]
\centering
\includegraphics[scale=0.2]{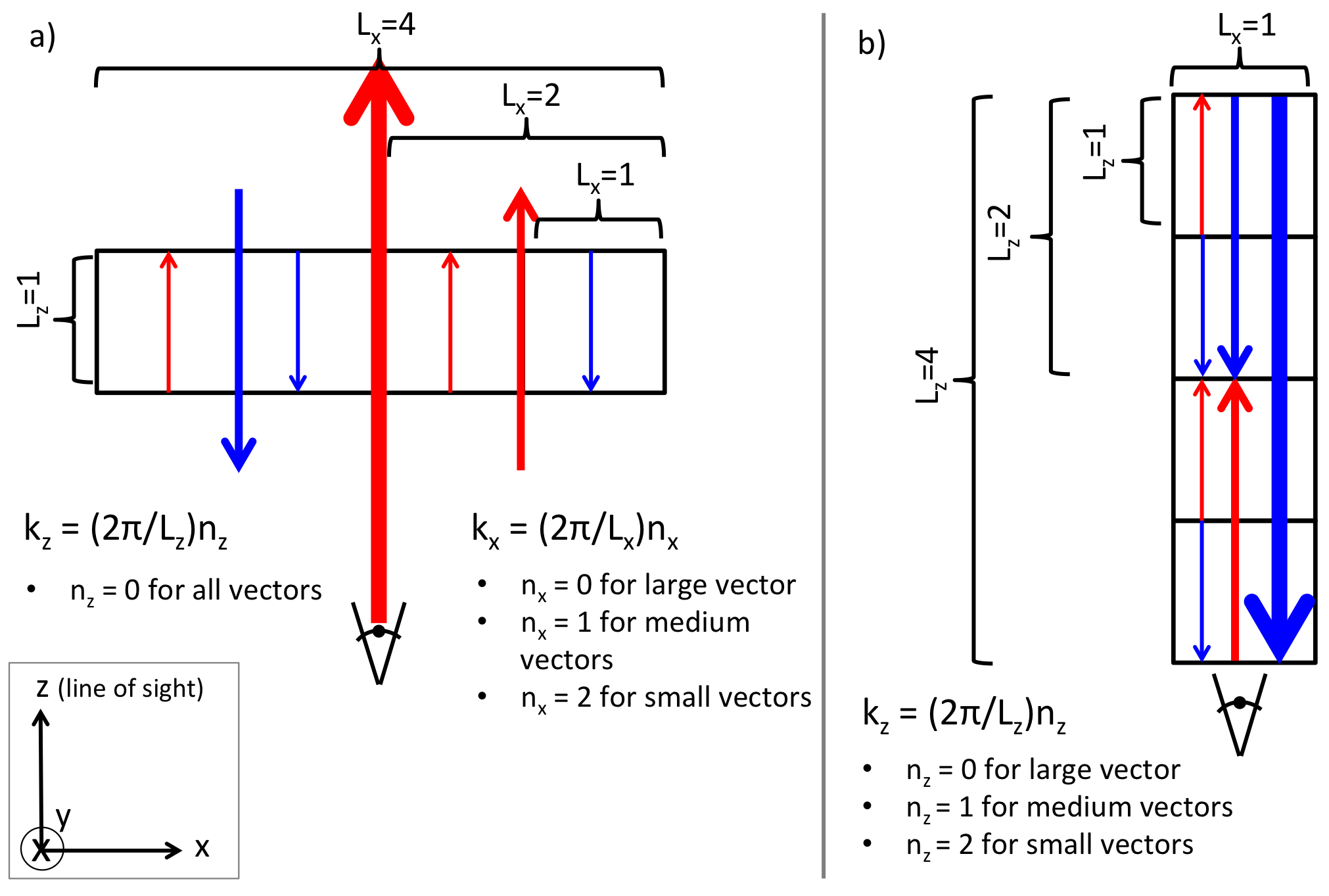}
\caption{ \small One-dimensional cartoons explaining our
  interpretation of the kinematics of spatially resolved dendrogram
  structures in the framework of isotropic turbulence. The spatial
  dimension of each structure is represented by four square cells,
  each one unit by one unit in area.  The colored vectors in each
  figure represent turbulent velocity components along the
  line-of-sight (blue = blue-shifted velocity component, red =
  red-shifted velocity component). The length of the velocity vectors
  are not to scale, but are simply used to show that there is larger
  velocity power on the larger scales, and vice-versa.  \textit{(a)}:
  We are looking across the resolved major axis ($x$-direction) of
  this filament, and down the minor axis ($z$-direction;
  line-of-sight).  The \vlsr{} of the filament at each resolved $L$=1
  cell is determined by the velocity of all vectors influencing the
  gas in that cell with $k_{z}=0$. Since all vectors have $k_{z}=0$,
  for example, the left-most cell has \vlsr{} contributions from a
  small red vector, a medium blue vector, and a large red vector,
  while the right-most cell has \vlsr{} contributions from a small
  blue vector, a medium red vector, and a large red vector.  The
  \vlsr{} variation ($\Delta$\vlsr) across the four resolved cells is
  the standard deviation of the \vlsr{} in each $L$=1 cell.  If this
  structure were half as wide, then $\Delta$\vlsr{} would decrease
  because each cell would only have contributions from a small and a
  medium vector.  The non-thermal velocity dispersion along each
  resolved line-of-sight ($\sigma_{\mathrm{nt}}$) is zero because
  there are no $k_{z} \ne 0$ vectors along the line of sight of each
  cell.  The total velocity dispersion is a combination of vector
  components with $k \ne 0$, which is determined by small and medium
  vectors across $x$-direction.  \textit{(b)}: We are looking down the
  major axis of this filament ($z$-direction; line-of-sight), through
  the single resolved spatial element in the $x$-direction. The
  \vlsr{} of the filament is determined by the velocity of the large
  vector because it is the only vector with $k_{z}=0$; oscillations of
  the small and medium vectors average each other out along the
  line-of-sight, and they do not contribute to the \vlsr{}. The
  \vlsr{} variation ($\Delta$\vlsr) is zero because there
  is only one resolved \vlsr{} value for this object. The non-thermal
  velocity dispersion ($\sigma_{\mathrm{nt}}$) along the single
  line-of-sight is determined by the small and medium vectors, since
  both have $k_{z} \ne 0$. If this structure were half as deep, then
  $\sigma_{\mathrm{nt}}$ would decrease because it would be determined
  by the small vectors alone. The total velocity dispersion is a
  combination of vector components with $k \ne 0$, which is determined
  by small and medium vectors along the $z$-direction in this case; it
  equals $\sigma_{\mathrm{nt}}$ since there is only one resolved
  line-of-sight.  }
\label{fig:cartoon}
\end{figure}

In this framework, the total linewidth of a structure is the sum of
components with non-zero $k_{x}$, $k_{y}$, or $k_{z}$. The resolved
linewidths ($\Delta$\vlsr{} and $\langle \sigma
\rangle_{\mathrm{nt}}$) are explained below with the aid of
Figure~\ref{fig:cartoon}.

The observed centroid velocity at a resolved projected position in the
structure, V$_{\mathrm{lsr}}(x,y)$, is the sum of components in
Equation~\ref{eq:turb} that have $k_{z}$=0. The V$_{\mathrm{lsr}}$
dependence on spatial scale via $v_{p}(k_{x},k_{y}; k_{z}=0)$ is
assumed to depend statistically only on $\vert {\bf k} \vert$ if the
turbulence is isotropic.  Therefore, \vlsr(x,y) has the same
statistical dependence on $\vert {\bf k} \vert = \sqrt{k_{x}^{2} +
  k_{y}^{2}}$ (which we can measure) that the underlying turbulent
velocity, $v_{z,\mathrm{turb}}(x,y,z)$, has on $\vert {\bf k}
\vert=\sqrt{k_{x}^{2} + k_{y}^{2} + k_{z}^{2}}$ (which we cannot
measure). The \vlsr(x,y) variation, $\Delta$\vlsr, which is the
standard deviation of V$_{\mathrm{lsr}}(x,y)$ over all resolved
positions in a structure, will scale with the $(x,y)$-size of the
structure according to the turbulent scaling relation of the cloud
(see \citet{1985ApJ...295..479D} for a discussion of spatial
  correlation properties of centroid velocity fields).  To visualize
this, we consider the structure on the left-side of
Figure~\ref{fig:cartoon}.  We call the full 4-cell structure,
structure A, and any contiguous 2-cell substructure within it,
structure B.  Structure A is wider than structure B in the $(x,y)$
directions on the sky, so the minimum, non-zero $k_{x}$ and $k_{y}$
must be smaller in structure A than structure B.  Assuming $v_{p}$ has
a (statistical) inverse power-law dependence on $\vert {\bf k} \vert$,
this implies higher power in structure A, with a greater contribution
to the V$_{\mathrm{lsr}}$ variation across the $(x,y)$ face of
structure A compared to structure B.

The non-thermal velocity dispersion at each resolved position in a
structure, $\sigma_{\mathrm{nt}}(x,y)$, is obtained from the sum of
turbulent components with all possible $k_{x}$ and $k_{y}$, and
non-zero $k_{z}$, along each line-of-sight. For a given $(x,y)$, the
total line-of-sight velocity is a sum: $v_z(z)=v_{z,\mathrm{turb}}(z)
+ v_{\mathrm{th}}(z)$, where $v_{\mathrm{th}}$ is the thermal
component of the velocity.  The mean value of $v_{\mathrm{th}}(z)$ is
zero, and the mean value of $v_{z,\mathrm{turb}}(z)$ is
V$_{\mathrm{lsr}}$.  Thus, assuming that $v_{z,\mathrm{turb}}(z)$ and
$v_{\mathrm{th}}(z)$ are uncorrelated, for a given line-of-sight, the
square of the total velocity dispersion, $\sigma^{2}(x,y)$, is:
\begin{equation}
\sigma^{2}(x,y) = \langle v_{z}^{2}\rangle - \mathrm{V}_{\mathrm{lsr}}^2 = \langle v_{z,\mathrm{turb}}^2\rangle
- \mathrm{V}_{\mathrm{lsr}}^2 + \langle v_{\mathrm{th}}^2 \rangle + 2 \langle
v_{z,\mathrm{turb}}\rangle \langle v_{\mathrm{th}}\rangle.
\label{eq:new}
\end{equation}
The last term in Equation~\ref{eq:new} is zero because the mean value
of $v_{\mathrm{th}}(z)$ is zero. The second-to-last term in
Equation~\ref{eq:new} is equivalent to the square of the thermal
velocity dispersion, $\sigma_{\mathrm{th}}^2$; $\sigma^{2} -
\sigma_{\mathrm{th}}^{2}$ is equal to the non-thermal component of the
velocity variance, $\sigma_{\mathrm{nt}}^{2}$, which means that the
first two terms Equation~\ref{eq:new} contribute to the non-thermal
velocity dispersion. The combination of the first two terms is
\begin{equation}
\langle \vert \sum_{k_{x},k_{y};k_{z}\ne 0} v_{p}(k_{x},k_{y},k_{z}) \exp (i {\bf k}\cdot {\bf x}) \vert^{2} \rangle,
\label{eq:new2}
\end{equation}
where the summation includes combinations with all $k_x$, $k_y$, and
with all $k_{z}$ except $k_{z}=0$; {\bf k} is the wavenumber vector,
and {\bf x} is the spatial vector.  
This shows that the mean
non-thermal velocity dispersion for our structures,
$\langle \sigma \rangle_{\mathrm{nt}}$, defined as the mean of
$\sigma_{\mathrm{nt}}(x,y)$ at each resolved position, will scale with
the $z$-depth of the structure according to the turbulent scaling
relation of the cloud.  Along any given line-of-sight, a larger
$L_{z}$ implies that
 the minimum nonzero $k_{z} = 2 \pi/L_{z}$ is
smaller, such that the contribution to Equation~\ref{eq:new2} would be
larger (under the assumption that $v_{p}({\bf k})$ statistically
increases with decreasing $\vert {\bf k}\vert$).  To visualize this,
we can consider the structure on the right-side of
Figure~\ref{fig:cartoon}.  We call the full 4-cell structure,
structure A, and any contiguous 2-cell substructure within it,
structure B. Structure A extends deeper into the sky than structure B,
so the minimum, non-zero $k_{z}$ in structure A is smaller than the
minimum, non-zero $k_{z}$ in structure B. This implies higher power in
structure A, with a larger contribution to $\sigma_{\mathrm{nt}}(x,y)$
in structure A compared to structure B.

Based on the arguments above, the scaling relation between
V$_{\mathrm{lsr}}$ variation and projected size, $l$, ($\Delta$\vlsr{}
$\propto l^{q}$) should have the same power-law dependence as the
scaling relation between mean non-thermal velocity dispersion and
depth, $d$, into the sky ($\langle \sigma \rangle_{\mathrm{nt}}
\propto d^{q}$) if the turbulence is isotropic and we are observing
all the gas along each line-of-sight.  But we saw in
Figure~\ref{fig:decon} that the dependences of $\Delta$\vlsr{} and
$\sigma_{nt}$ on projected structure size are very different. The
simplest explanation of this difference is that the projected size of
an object need not be the same as its line-of-sight depth into the
sky. The size axis in Figure~\ref{fig:decon} (right) is an estimate
based on the geometric mean of the projected size.  If every object,
no matter its projected size, has the same depth, then we should
measure a similar non-thermal velocity dispersion for those
objects. Therefore, we interpret the shallow dependence of $\langle
\sigma \rangle_{\mathrm{nt}}$ with size as an indication that our
collection of dendrogram structures have similar depths.

\begin{figure}[]
\centering
\includegraphics[scale=1.0]{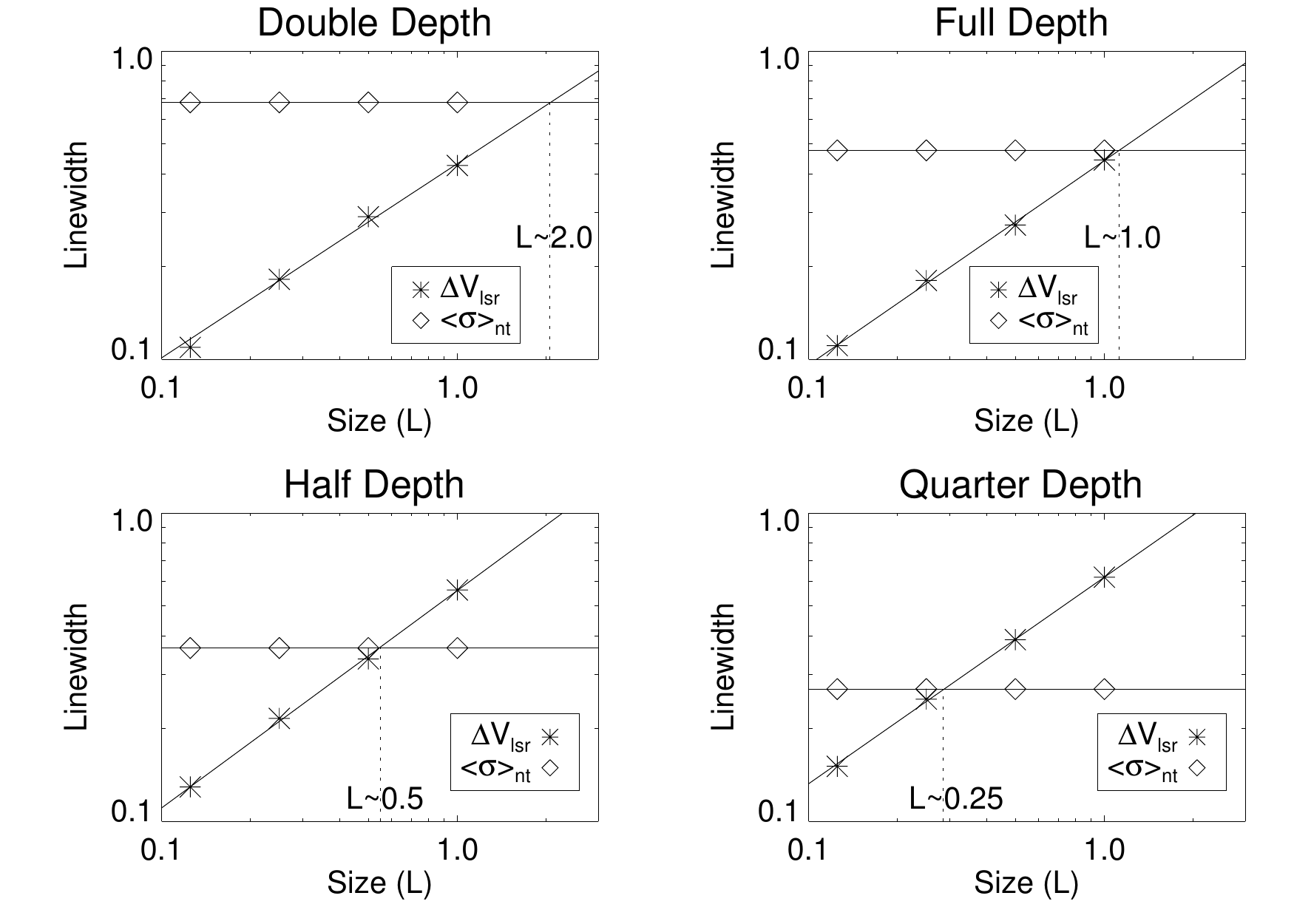}
\caption{ \small Size-linewidth relations from a numerical
    realization of an isotropic turbulent power spectrum. The four
    plots represent results from four different boxes taken from the
    realization: the ``double depth'' box is twice as deep as it is
    wide (L$\times$L$\times$2L), the ``full depth'' box is cubic
    (L$\times$L$\times$L), the ``half depth'' box is half as deep as
    it is wide (L$\times$L$\times$L/2), etc. The vertical axis
    represents the two linewidths we discussed in text: $\langle
    \sigma \rangle_{\mathrm{nt}}$ and $\Delta$\vlsr{} (in arbitrary
    units). The horizontal axis represents the size-scale we sampled
    across the face of the box when calculating the linewidths: a
    size-scale of 1 represents sampling across the full L$\times$L
    scale, a size-scale of 0.5 represents sampling across
    smaller-scale L/2$\times$L/2 segments, etc. There are several
    things to notice in these panels: 1) $\langle \sigma
    \rangle_{\mathrm{nt}}$ is independent of size, and its magnitude
    decreases by $\sqrt{2}$ as the box depth is halved from panel to
    panel, 2) $\Delta$\vlsr{} increases with increasing size, 3) The
    magnitude of $\langle \sigma \rangle_{\mathrm{nt}}$ and
    $\Delta$\vlsr{} cross near the size that represents the depth of the
    box.}
\label{fig:newsim}
\end{figure}

To illustrate these effects, we created a numerical realization of a
three-dimensional, isotropic turbulent power spectrum, with power
spectrum scaling $v^{2} \propto k^{-4}$. Using an arbitrary length
scale, L, the position-position-position box had $x,y,z$ dimensions of
L$\times$L$\times$2L, where L corresponds to 512 pixels.  We used this
realization along with three sub-boxes with different L$_{\mathrm{z}}$
to demonstrate how $\langle \sigma \rangle_{\mathrm{nt}}$ and
$\Delta$\vlsr{} depend on structure depth for a single turbulent
realization. The original box had dimensions of L$\times$L$\times$2L
(the ``double depth'' box), the first sub-box was L$\times$L$\times$L
(the ``full depth'' box), the second was L$\times$L$\times$L/2 (the
``half depth'' box), and the third was L$\times$L$\times$L/4 (the
``quarter depth'' box). For each of these sub-boxes (with uniform
density), we created PPV cubes and then processed them similar to the
observations. The end products were centroid velocity and velocity
dispersion maps across the L$\times$L $x,y$ surface of each numerical
box. We segmented each L$\times$L surface into equally sized square
regions with side L, L/2, L/4, L/8, and calculated $\Delta$\vlsr{} and
$\langle \sigma \rangle_{\mathrm{nt}}$ within each segment to show how
they scale with size. The results for each box are shown in
Figure~\ref{fig:newsim}. Evidently, $\langle \sigma
\rangle_{\mathrm{nt}}$ has no dependence on size, while
$\Delta$\vlsr{} increases with size, similar to the behavior seen in
the observations in Figure~\ref{fig:decon}. The size where
$\Delta$\vlsr{} crosses $\langle \sigma \rangle_{\mathrm{nt}}$
corresponds well with the depth of the numerical cube.

If turbulence in the observed region has similar behavior to that in
the isotropic turbulent realization, then we can use the correlation
of \vlsr{} variation with size to estimate the depths of the Barnard~1
structures. For this correlation, we know the projected size is
directly related to $\Delta$\vlsr, so there is no size ambiguity as
there is in the correlation between projected size and $\langle \sigma
\rangle_{\mathrm{nt}}$. To estimate the typical depth of the cloud, we
can find the size scale at which the best-fit line for $\Delta$\vlsr{}
versus projected size is equal to the best-fit line for $\langle
\sigma \rangle_{\mathrm{nt}}$ versus projected size. This corresponds
to $\sim$0.11~pc for the MCZ, and $\sim$0.18~pc for the RSWZ. Under
the assumption that turbulence is isotropic, this implies that
Barnard~1 is on the order of 0.1-0.2~pc deep, comparable to the
largest projected width ($\sim$0.2-0.3~pc) of individual structures.

Since the largest-scale dust structure in Barnard~1 extends over 1~pc
on the sky, the overall region may be more flattened than spherical at
the largest scales.  Numerical simulations over the past 15 years have
consistently shown that high-density sheets and filaments are a
generic result of strongly supersonic turbulence with parameters
comparable to those in observed clouds \citep[e.g., reviews
  by][]{2004ARA&A..42..211E, 2004RvMP...76..125M,
  2007ARAA..45..565M}. But measuring the depth of an individual
structure in a molecular cloud is a difficult task.  Comparison to
realizations of turbulence suggests that the high angular resolution,
large area observations from CLASSy allow us to estimate the physical
depths of individual structures in molecular clouds, thus providing
new observational insight into the true nature of those
clouds\footnote{We emphasize, however, that it is important to confirm
  the behavior seen in simple, isotropic turbulent realizations with
  fully-realistic turbulent simulations of clouds.  For best
  comparison to observations, it will be valuable to create synthetic
  PPV data cubes via radiative transfer modeling, rather than assuming
  uniform excitation and optically thin conditions.}.  Our
observational result suggests that Barnard~1 is an over-dense region
within the larger Perseus Molecular Cloud that could have formed a
sheet-like geometry from supersonic turbulence. Lee et. al (2014,
submitted) present the same size-linewidth relations using CLASSy
observations of Serpens Main, and find similar behavior.  We will
compare the results from all our CLASSy regions in an upcoming
cross-comparison paper.

\section{SUMMARY}
  
We presented the data release for the Barnard~1 region of the CARMA
Large Area Star Formation Survey. CLASSy spectrally imaged over 800
total square arcminutes of the Perseus and Serpens Molecular Clouds at
7\arcsec{} angular resolution.

\begin{enumerate}

\item We spectrally imaged \NtwoH, HCN, and \HCO{} ($J=1\rightarrow0$)
  emission across 150 square arcminutes of Barnard~1. The final data
  products are position-position-velocity cubes with full line
  emission recovery obtained through joint deconvolution with
  single-dish observations. The velocity resolution of the data is
  $\sim$0.16~\kms.

\item Four compact continuum sources were detected at $>$5-$\sigma$ at
  3~mm, all in the main core zone. B1-c, B1-b South, and B1-b North
  were previously known; we report a new detection of compact emission
  towards the Per-emb~30 continuum source.

\item The \NtwoH{} ($J=1\rightarrow0$) gas morphology closely matches
  dust continuum observations of \textit{Herschel}, while HCN and
  \HCO{} ($J=1\rightarrow0$) emissions are weaker throughout most of
  the field and show less correlation with the long wavelength dust
  emission. HCN and \HCO{} also well-trace outflows in the main core
  zone.

\item Spectral line fitting of the molecular line data shows that the
  Barnard~1 main core is much more kinematically complex than the
  filaments and clumps that extend to its southeast; these filaments
  and clumps are characterized by more uniform centroid velocities and
  lower velocity dispersions.

\item We used dendrograms to identify \NtwoH{} gas structures in
  Barnard~1. The motivation for using dendrograms instead of a more
  traditional clumpfinding algorithm was the need to analyze the
  morphological and kinematic structure of dense gas across the wide
  range of spatial scales captured in our CLASSy data. We found that
  dendrograms are better able to quantify that range of spatial
  scales.  We created a new, non-binary adaptation to the standard,
  binary dendrogram algorithm to ensure that the dendrograms represent
  the true hierarchy of the emission within the noise limits of real
  data, and that tree statistics can be used to quantify that
  hierarchy.

\item The non-binary dendrogram of Barnard~1 contains 41 leaves and 13
  branches. We calculated three simple tree statistics using the
  dendrogram: the maximum branching level, the mean path length, and
  the mean branching ratio.  The tree structure representing the dense
  gas around the main core is the most complex, with four hierarchy
  levels and the highest contrast leaves.  The tree statistics give
  insight into the type and amount of fragmentation a region has
  undergone, and will be used to compare the hierarchical complexity
  of the different CLASSy regions.

\item We characterized the spatial properties of the dendrogram
  structures and derived structure sizes ranging from $\sim$0.01 to
  0.34~pc. The high angular resolution data reveal a variety of
  irregular shapes, showing that star-forming gas is not composed of
  well ordered spheroids and filaments on the smallest scales. We also
  characterized the kinematic properties of the structures and found
  that, in general, branches have larger \vlsr{} variation, but
  similar mean velocity dispersion, compared to the leaves. The gas
  surrounding the most massive compact continuum cores have the
  largest velocity dispersions in the entire region.

\item Using the spatial and kinematic properties of the dendrogram
  leaves and branches, we estimated the depth of the Barnard~1 cloud
  to be $\sim$0.1-0.2~pc. This estimate was made by comparing two
  size-linewidth relations: one using the mean non-thermal velocity
  dispersion of the dendrogram objects, which is sensitive to the
  depth of the cloud, and the other using the \vlsr{} variation of the
  objects, which is sensitive to the projected size of the cloud.  The
  mean non-thermal velocity dispersion varied very little with
  structure size, while the \vlsr{} variation varied steeply with
  size. We interpreted this as an indication that Barnard~1 is more
  flattened than spherical on the largest scales. This method is
  a powerful tool for observationally probing the structure of
  molecular clouds into the plane of the sky.
\end{enumerate}

 The science-ready spectral line data cubes for the Barnard~1 region
 can be found through the online version of Figure~5. We will also
 host data products at our CLASSy website:
 http://carma.astro.umd.edu/classy. We welcome the community to make
 use of the data.

\acknowledgments The authors would like to thank referee Alyssa
Goodman for encouraging critiques that improved the paper, and all
members of the CARMA staff that made these observations
possible. CLASSy was supported by AST-1139990 (University of Maryland)
and AST-1139950 (University of Illinois).  Support for CARMA
construction was derived from the Gordon and Betty Moore Foundation,
the Kenneth T. and Eileen L. Norris Foundation, the James S. McDonnell
Foundation, the Associates of the California Institute of Technology,
the University of Chicago, the states of Illinois, California, and
Maryland, and the National Science Foundation. Ongoing CARMA
development and operations are supported by the National Science
Foundation under a cooperative agreement, and by the CARMA partner
universities.

 {\it Facility:} \facility{CARMA}%

\bibliography{mybib_b1}

\appendix
\section{COMPARING CLUMP-FINDING AND DENDROGRAMS FOR CLASSy DATA}

\begin{figure}[H]
\centering
\includegraphics[scale=0.5]{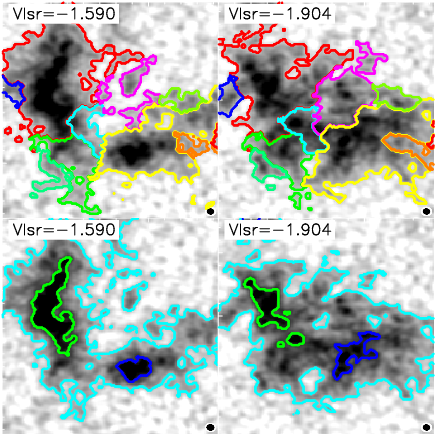}
\caption{\footnotesize Two \NtwoH{} channel maps of the B1-b region
  for a comparison of Cloudprops (top) and dendrogram (bottom) object
  identification results. Each Cloudprops and dendrogram object is
  contoured with a unique color---this particular region is broken up
  into nine distinct Cloudprops objects, and three dendrogram
  objects. Cloudprops tends to put all emission into smaller-scale,
  independent clumps, while dendrograms identifies peaks of emission
  as leaves (green and blue contours in the online version), and then
  identifies lower intensity contour(s) that surround them as branches
  (cyan contour in the online version). We argue that the dendrogram
  approach is the more appropriate one for high angular resolution
  studies of nearby molecular clouds, since it captures the
  hierarchical nature of the gas emission across the wide range of
  spatial scales that exists in those regions.}
\label{fig:cloudvdendroIRAS2A}
\end{figure}

 We use the Cloudprops algorithm \citep{2006PASP..118..590R} and an
 IDL dendrogram algorithm \citep{2008ApJ...679.1338R} to identify
 \NtwoH{} emission structures in our CLASSy data cubes.  The goal is
 to compare clump-finding and dendrogram object identification methods
 when applied to nearby star forming regions observed with high
 angular resolution.  Both algorithms analyze the emission in
 position-position-velocity cube to segment the emission into
 structures.  Figure~\ref{fig:cloudvdendroIRAS2A} shows Cloudprops
 (top) and dendrogram (bottom) contours for identified structures,
 overplotted on two \NtwoH{} velocity channels around the B1-b
 continuum cores. The common color contours represent the same
 structures in the left and right panels. There are clearly major
 differences in the number and shapes of the identified structures
 between the two methods.

 We find that the Cloudprops algorithm forces emission into
 small-scale clumps, even when the data does not show clear regions
 with strong intensity enhancements. When visually inspecting the
 emission channels in Figure~\ref{fig:cloudvdendroIRAS2A}, there is no
 physical reason to think that the plateau of emission west of the
 B1-b cores is made up on several, independent clumps of dense gas
 that are separated with orderly borders.  The technical reason for
 the clumpy breakdown is that Cloudprops first locates the highest
 level, local closed isocontours. Next, the algorithm steps down in
 intensity dividing the lower intensity emission between the stronger
 peaks according to a nearest-neighbor algorithm. This division of
 lower intensity emission across a fairly uniform emission plateau
 leads to the arbitrary borders seen in the top row of
 Figure~\ref{fig:cloudvdendroIRAS2A}. Cloudprops and other
 clump-finding algorithms work best for sparse fields that have
 resolved separations between objects---identifying giant molecular
 clouds in an extragalactic source is one example of where they work
 well. But attempting to decompose nearby, well-resolved, molecular
 gas emission into many small clumps with fairly straight borders does
 not have good physical motivation. Similar conclusions were found in
 \citet{2009Natur.457...63G}, where the authors showed that CLUMPFIND
 \citep{1994ApJ...428..693W} makes artificial attempts to fill in the
 structure between meaningful clumps.
 
 Instead of partitioning all of the emission into individual clumps
 according to a nearest-neighbor scheme, it makes more physical sense
 to interpret it as lower intensity structure surrounding the peaks
 within it. For example, the peaks could represent gas that fragmented
 from the lower intensity emission surrounding them. This scenario
 considers that small-scale features can form from existing
 large-scales features, and is a hierarchical way of thinking about
 the data. If we accept that molecular clouds are hierarchical, then
 this is a more physical way to interpret the emission, and it
 requires an algorithm that can track the emission structure as a
 function of contour level intensity---a dendrogram algorithm does
 just that.

\begin{figure}[H]
\centering
\includegraphics[scale=0.2]{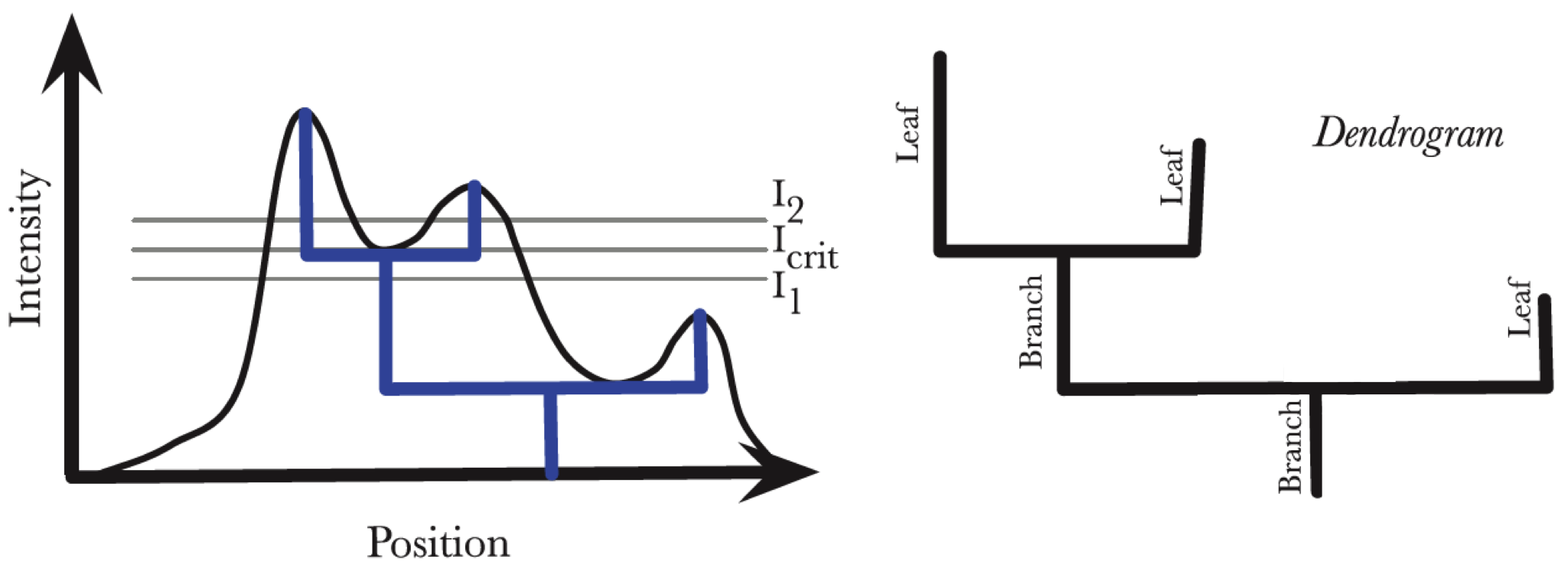}
\caption{\footnotesize This figure is taken from
  \citet{2008ApJ...679.1338R}.  The left image shows a
  one-dimensional, three-peak emission profile with the dendrogram
  representation as an inset. The dendrogram is reproduced on the
  right with the tree features labeled (the labels differ slightly
  from \citet{2008ApJ...679.1338R} to accommodate the naming
  conventions used in this paper). The dendrogram is formed by keeping
  track of the emission structure at a function of intensity
  level. For example, contouring the emission profile at the I$_{1}$
  level produces a single object, while contouring at I$_{2}$ produces
  two objects---the I$_{\mathrm{crit}}$ contour represents the level where the
  two strongest peaks in the emission profile merge together. In the
  dendrogram, this contour level is represented by a horizontal line
  connecting the two strongest leaves into the branch. Readers should
  see the original paper for a detailed discussion on dendrograms in
  all three dimensions.}
\label{fig:dendro1D}
\end{figure}

 How does a dendrogram algorithm work? We quickly review the basic
 dendrogram procedure using a one-dimensional emission profile seen in
 Figure~\ref{fig:dendro1D} \cite[from][]{2008ApJ...679.1338R}. (See
 that paper for more discussion of this example and the extension to
 two- and three-dimensions.) A one-dimensional, horizontal contour can
 be started at the absolute maximum of the emission profile and stepped
 downward to the base of emission. During this process, the peaks and
 merge levels of the three local maxima are identified and represented
 in a dendrogram (seen inset within the emission profile and repeated
 on the right with labels).

 A dendrogram is composed of leaves and branches. In our example, the
 three leaves of the dendrogram correspond to the three local maxima
 in the emission profile that do not break up into any
 substructure. The peak intensity of a leaf represents the peak
 intensity of its corresponding local maximum, while the lowest
 intensity of a leaf represents the intensity where the corresponding
 local maximum merges with another local maximum.  The two branches of
 the dendrogram correspond to lower-intensity emission that breaks up
 into substructure at higher intensities.  The upper branch represents
 the lower-intensity emission that seeds the two strongest leaves, and
 the lower branch represents even lower-intensity emission that
 encompasses all of the emission peaks.  The peak intensity of a
 branch represents the intensity where the leaves above it merge
 together, while the lowest intensity of a branch represents the
 intensity where it merges with another leaf or branch (the case of
 the left branch) or where it reaches the lowest measured intensity
 (the case of the lower branch). This algorithm can be extended from
 one-dimension to two- and three-dimensions to operate on
 position-position or position-position-velocity datasets. All of our
 object identification for CLASSy clouds was done in three-dimensions,
 even if we occasionally represent the identified objects as
 two-dimensional contours for visualization simplicity.

 Going back to our example in Figure~\ref{fig:cloudvdendroIRAS2A}, the
 dendrogram identified objects in the two \NtwoH{} channels are shown
 on the bottom row of the figure. Two peaks of emission (leaves) are
 identified with isolated contours (green and blue contours), and then
 get joined at a lower intensity level within a larger scale contour
 (a branch shown as the cyan contour).

 To summarize, the dendrogram identification method is more
 appropriate for our data and science goals. Our data capture
 small-scale dense gas structure due to the high-angular resolution of
 the interferometer, and they capture large-scale dense gas features
 because of the large-area mosaicing and full emission reconstruction
 with single-dish data.  Our science goals involve studying how
 small-scale gas features connect to the large-scale cloud
 features. The dendrogram approach lets us investigate the physical
 and kinematic properties of the dense gas across the wide range of
 spatial scales probed by our CLASSy data in a way that purely
 small-scale, clumpfind-like segmentation would not allow.

\section{A NEW NON-BINARY DENDROGRAM PROCEDURE}
 The standard dendrogram algorithm forms a purely binary tree --
 thinking about a tree growing from the base upward, every branch
 terminates in either two leaves, a leaf and a branch, or two
 branches. A single branch is not allowed to directly sprout three
 leaves, even if the leaves merge into the branch at exactly the same
 level. In this three leaf example, two of the leaves would get merged
 into a branch, and then that branch would merge with the remaining
 leaf into another branch. We refer to this as ``phantom branching''
 because an artificial branch was introduced only to enforce a binary
 merger requirement, as opposed to being motivated by the true nature
 of the data being modeled. This makes interpreting the true
 hierarchical nature of the data difficult, and makes a quantitative
 assessment of the dendrogram using tree statistics impossible (see
 \citet{1992ApJ...393..172H} for the use of tree statistics to
 interpret the hierarchical nature of emission). Going back to our
 example, we set out to allow all three leaves to merge into a single
 branch---we set out to allow non-binary mergers.

 Non-binary dendrograms are useful for two reasons: 1) they provide a
 collection of branches that respect the noise inherent in real data,
 and 2) they allow the user to quantitatively compare the hierarchical
 complexity of different emission regions, and correlate that
 complexity with properties of those regions (e.g., star formation
 efficiency, column density, etc.).

 The three-leaf example we used to motivate non-binary dendrograms is
 artificial; it is extremely unlikely for a real dataset to have three
 leaves merging at exactly the same intensity.  Typical behavior seen
 in real data sets may have two leaves merge into a branch at the
 1.0~\Jybm{} level, and then merge with a third leaf into another
 branch at the 0.98~\Jybm{} level.  Our key argument is that these
 three leaves should merge into a single branch, \textit{if} the
 sensitivity of the dataset is lower than the branching
 difference. For example, the sensitivity of our Barnard 1 dataset is
 0.13~\Jybm, and the standard dendrogram algorithm produces branching
 in increments less than 0.13~\Jybm{} to ensure binary mergers ---this
 produces a collection of branches that is not meaningful within the
 noise limits of the data, and limits the quantitative interpretation
 of the hierarchy.  Our new method alleviates the binary merger
 requirement and allows branches to sprout an unrestricted amount of
 leaves and/or branches.

 The key technical differences between our new non-binary method and
 the standard method are: 1) we restrict branching to intensity steps
 equal to integer values of the 1-$\sigma$ sensitivity of the data,
 instead of allowing branching at infinitely small intensity steps, 2)
 we have an algorithm that can cluster more than two objects into a
 single group, instead of being restricted to clustering two objects
 at a time. Another way to think about the difference is that the
 standard dendrogram code yields the one emission hierarchy that
 represents the one realization of the noise applied to the true
 emission being observed. Our modified dendrogram code instead
 represents an observable emission hierarchy within the noise limits
 of the data.

\begin{figure}[H]
\centering \includegraphics[scale=0.16]{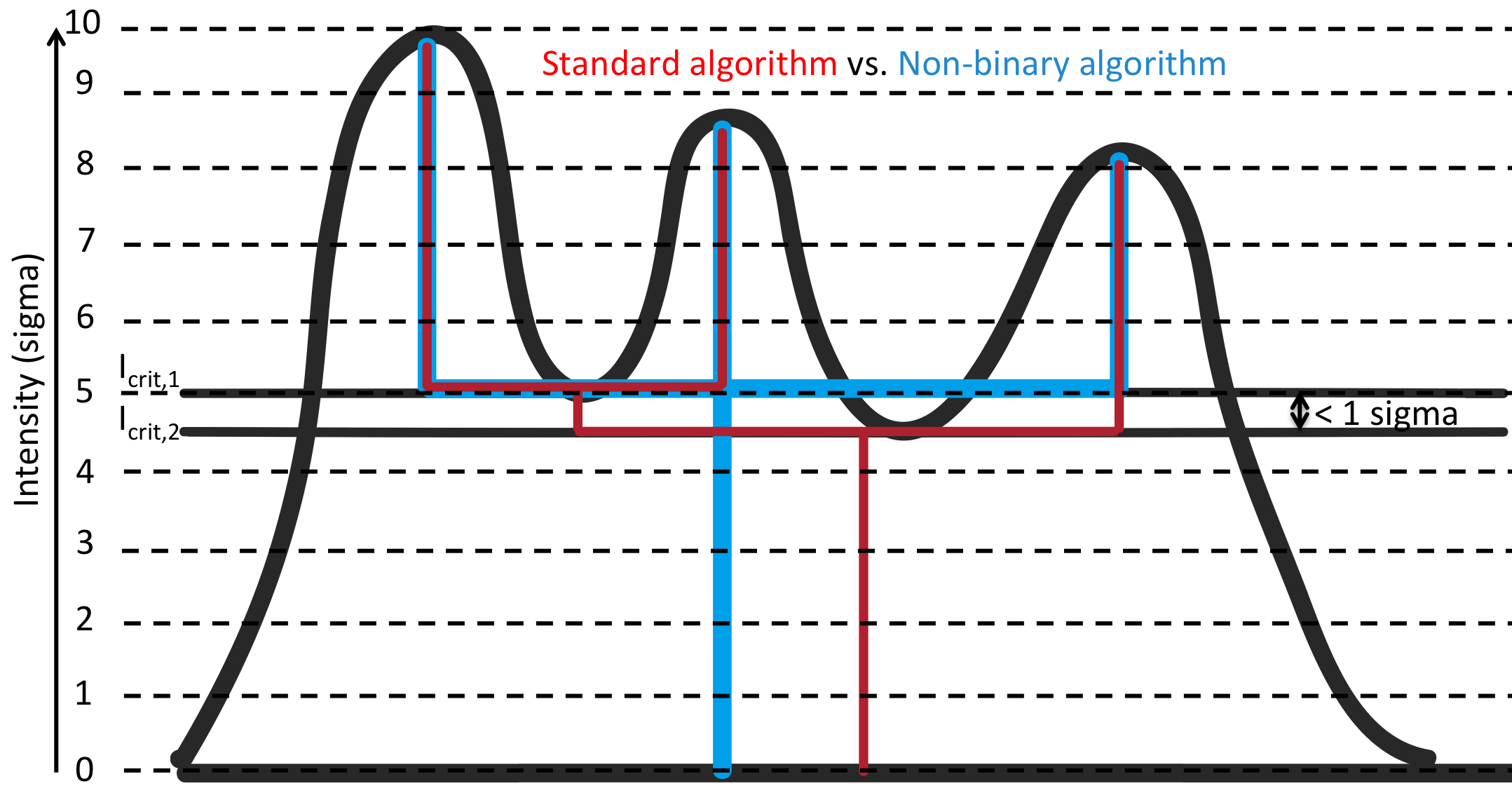}
\caption{\footnotesize Cartoon example of the difference between the
  standard dendrogram algorithm, and our new non-binary algorithm. The
  standard algorithm (red tree in the online version) forces binary
  mergers of dendrogram leaves and branches---the left and center
  peaks are first merged into a branch at the I$_{\mathrm{crit},1}$
  contour level, and that branch is then merged with the right peak at
  the I$_{\mathrm{crit},2}$ contour level. Since the two merger levels
  are separated by less than the 1-sigma sensitivity of the data, we
  argue that the branch between I$_{\mathrm{crit},1}$ and
  I$_{\mathrm{crit},1}$ is not motivated by the data, but by the
  requirement of binary mergers. Our new non-binary algorithm (cyan in
  the online version) merges all three peaks at the highest merger
  level, and assigns the remaining emission between
  I$_{\mathrm{crit},1}$ and I$_{\mathrm{crit},2}$ to a single
  branch. This represents an observable emission hierarchy within the
  noise limits of the data.}
\label{fig:dendroexample}
\end{figure}

A cartoon example of the difference is shown in
Figure~\ref{fig:dendroexample}. The cartoon shows a one-dimensional
intensity profile, where the intensity axis is quoted in terms of the
sigma sensitivity units of the data. Under the standard algorithm (red
tree), the two leftmost peaks merge together at the level of the first
saddle point (I$_{\mathrm{crit},1}$, which represents an intensity
level of 5-sigma) into a branch. The algorithm then merges that first
branch with the rightmost peak at the level of the second saddle point
(I$_{\mathrm{crit},2}$, which is less than 1-sigma lower than
I$_{\mathrm{crit},1}$) into a second branch. Our algorithm (cyan tree)
merges all three peaks into a single branch at the 5-sigma level since
the two saddle points are separated by less than the 1-sigma
sensitivity of the data.
 
Our algorithm comes with one obvious caveat. The resulting dendrogram
depends on the 1-$\sigma$ contouring of the data; if branching levels
are shifted slightly higher or lower, then the dendrogram can
change. For example, consider a dataset in Kelvin units (for visual
simplicity), with 0.10~K sensitivity, two leaves that peak at 4.00~K
and merge together at 1.00~K, and a third leaf that peaks at 3.00~K
and merges with the other leaves at 0.98~K. Our algorithm would set
1-$\sigma$ branching levels in 0.10~K steps starting from the 4.00~K
peak, which means that 1.00~K is an available merge level. All three
leaves would be merged into a single branch at 1.00~K using our
non-binary algorithm. But if the peak intensity is 3.98~K instead of
4.00~K, then 1.00~K is not an available branching level anymore -- but
1.08~K and 0.98~K are. For that case, the two strongest leaves would
merge at 1.08~K, and then merge with the third leaf at 0.98~K. This
simple example shows how our dendrograms, and the tree statistics
derived from them, can be slightly altered by shifts in the 1-$\sigma$
contouring of the data. We went from a dendrogram with one branching
level, a branching ratio of three, and a mean path length of one, to a
dendrogram with to two branching levels, a mean branching ratio of
two, and a mean path length of 1.66.

This caveat produces only small changes in the dendrogram and tree
statistics when tested on the actual Barnard~1 data presented in the
paper. We shifted the 1-$\sigma$ branching levels by 0.33-$\sigma$;
compared to the results presented in section~6, the mean branching
ratio changed from 3.9 to 3.8, and the mean path length and maximum
branching level were unchanged.  We argue that the benefits of this
new non-binary algorithm outweigh this caveat, particularly when
science goals include: 1) correlating the physical properties of
several star forming regions with differences in dendrogram structure
(by comparing tree statistics of different regions, which is not
possible with the standard algorithm), and 2) comparing the structure
and kinematics of large-scale and small-scale structures within a
single region (by comparing the spatial and kinematic properties of
leaves and branches, which is more difficult for the end-user if the
branch list is contaminated by phantom branching from the standard
algorithm).

\end{document}